\documentclass[useAMS,usenatbib]{mn2e}
\usepackage{epsfig}

\defcitealias{2000MNRAS.316..357G}{GdP00}
\defcitealias{1999ApJS..123....3L}{SB99}
\defcitealias{2000ApJ...539..718C}{CF00}
\defcitealias{2000ApJ...533..682C}{CALZ00}
\defcitealias{2002MNRASnotyetII}{Paper II}
\newcommand{\gil}{\citetalias{2000MNRAS.316..357G}}
\newcommand{\SB}{\citetalias{1999ApJS..123....3L}}
\newcommand{\cf}{\citetalias{2000ApJ...539..718C}}
\newcommand{\calz}{\citetalias{2000ApJ...533..682C}}
\newcommand{\pii}{\citetalias{2002MNRASnotyetII}}

\title{Stellar populations in local star-forming galaxies.\\I.--Data 
and modelling procedure.}
\author[P.~G.~P\'{e}rez-Gonz\'{a}lez et al.]
       {P.~G.~P\'{e}rez-Gonz\'{a}lez$^{1}$, A.~Gil de
       Paz,$^{5,2,}$$^{1}$ J.~Zamorano,$^{1}$ J.~Gallego,$^{1}$
\newauthor
A.~Alonso-Herrero$^{3}$ and A.~Arag\'{o}n-Salamanca$^{4}$\\
$^{1}$Departamento de Astrof\'{\i}sica, Facultad de F\'{\i}sicas, Universidad Complutense, E-28040 Madrid, Spain\\
$^{2}$NASA/IPAC Extragalactic Database, California Institute of Technology, MS 100-22, Pasadena, CA 91125, USA\\
$^{3}$Steward Observatory, The University of Arizona, Tucson AZ 85721, USA\\
$^{4}$School of Physics and Astronomy, University of Nottingham, NG7 2RD, England\\
$^{5}$current address:  The Observatories of the Carnegie Institution of Washington, 813 Santa Barbara St., Pasadena, CA 91101, USA}
\date{Received \today}

\pagerange{\pageref{firstpage}--\pageref{lastpage}}
\pubyear{2002}

\begin{document}

\maketitle

\label{firstpage}

\begin{abstract}

We present an analysis of the integrated properties of the stellar
populations in the {\it Universidad Complutense de Madrid} (UCM)
Survey of $\mathrm H\alpha$-selected galaxies. In this paper, the
first of a series, we describe in detail the techniques developed to
model star-forming galaxies using a mixture of stellar populations,
and taking into account the observational uncertainties. We assume a
recent burst of star formation superimposed on a more evolved
population. The effects of the nebular continuum, line emission and
dust attenuation are taken into account. We also test different model
assumptions including the choice of specific evolutionary synthesis
model, initial mass function, star formation scenario and the
treatment of dust extinction.  Quantitative tests are applied to
determine how well these models fit our multi-wavelength observations
for the UCM sample. Our observations span the optical and near
infrared, including both photometric and spectroscopic data. Our
results indicate that extinction plays a key role in this kind of
studies, revealing that low- and high-obscured objects may require
very different extinction laws and must be treated differently. We
also demonstrate that the UCM Survey galaxies are best described by a
short burst of star formation occurring within a quiescent galaxy,
rather than by continuous star formation. A detailed discussion on the
inferred parameters, such as the age, burst strength, metallicity,
star formation rate, extinction and total stellar mass for individual
objects, is presented in paper II of this series.


\end{abstract}


\section{Introduction}

One of the main issues in today's Astrophysics is how present day
galaxies formed and how they have evolved over time. A considerable
observational effort is being made to study galaxies from the earliest
possible times to the present. Our knowledge of the faint galaxy
populations over the $0<z<5$ range has experienced remarkable progress
in a relatively short period of time (see the reviews by
\citealt{1997ARA&A..35..389E} and
\citealt{2000ARA&A..38..667F}). One of the main aims of these studies is to
find the  progenitors of the local galaxy population. While the majority of 
local galaxies seem to fit reasonably well into the Hubble sequence, this
morphological classification scheme breaks down at surprisingly low redshifts
\citep[$z\sim0.3$-$0.5$; see][and  references therein]{2002dgkdp}. 
Moreover, new classes of distant objects have been discovered, such as
{\it Ultraluminous Infrared Galaxies},
\citep{1983ApJ...269..352S}, {\it Extremely Red Objects},
\citep{2000AJ....120..575Y} as well as bright UV galaxies \citep[{\it Lyman
Break Galaxies}, LBGs,][] {1996ApJ...462L..17S,1999ApJ...519....1S}. The
luminosity of these objects, both the reddest and the bluest, is mainly
dominated by massive knots of newly-formed stars (starbursts), with different
amounts of dust extinction.

A complementary approach to understand how present-day galaxies came
into being is to study in detail the properties of local galaxies, and
in particular their star-formation histories.  In this respect, it is
important to quantify the relative importance of the current episode
of star formation in comparison to the underlying older stellar
populations. Indeed, even in high redshift objects, stars formed
before the currently-observed star formation episode must have been
present in order to produce the observed metal and dust content.
Examples of such high-$z$ objects include SCUBA sources
\citep{1998Natur.394..241H} and LBGs (see
\citealt{2001PASP..113.1449C} and references therein). Moreover, an accurate
determination of the total stellar mass in both local and distant galaxies is a
necessary step towards understanding their formation (see, e.g., 
\citealt{1998ApJ...508..539P,2001ApJ...554..981P}).  Our group is actively
working on the detailed study of galaxies in the Local Universe so that their
properties can be compared with distant ones.  The  techniques developed and
tested with local galaxies  will have direct application in high-{\it z}
studies.

The {\it Universidad Complutense de Madrid\/} (UCM) Survey was carried out in
order to  perform a comprehensive study of star-forming galaxies in the Local
Universe (\citealt{1994ApJS...95..387Z,1996ApJS..105..343Z}; see also 
\citealt{1999ApJS..122..415A}).  This $\mathrm H\alpha$-selected galaxy sample has been
extensively studied at optical and near infrared wavelengths (see next
section). It has also been used to determine the local $\mathrm H\alpha$
luminosity function and star formation rate density
\citep{1995ApJ...455L...1G}, providing a low-$z$ benchmark for
intermediate and high-{\it z} studies \citep[see, for
example,][]{2000PASJ...52...73I,2000A&A...362....9M,2000A&A...362..509V,astro-ph/0111390}.
Recently, the UCM Survey is being extended to higher redshifts
\citep{2001A&A...379..798P}.

The present series of papers aims at determining the main properties
of the stellar populations in the UCM Survey galaxies, accounting both
for the newly-formed stars and the underlying evolved populations. We
make use of the extensive multi-wavelength data available for the
sample. A direct precursor of the current study was presented in
\citet[hereafter \gil]{2000MNRAS.316..357G}, where we characterized
the stellar content of a smaller subsample of 67 UCM galaxies,
constraining their ages, metallicities and relative strength of the
current star-formation episode.  A sophisticated statistical technique
was developed by \gil\ to compare measurements and model
predictions. We now present results for virtually all the UCM
galaxies, increasing the sample by a factor of 2.5. We have also
included additional photometry in the $B$-band
\citep{2000A&AS..141..409P}, and use a more elaborated spectral
synthesis method.

The present paper will focus on the modelling technique and the
observational data used to test it.  We will discuss the model input
parameters that best describe the observed properties of the UCM
galaxies, including the initial mass function (IMF), star formation
scenarios and extinction prescriptions. We will also study in detail
how well our modelling techniques are able to reproduce the
observations.  In \citet[\pii\, hereafter]{2002MNRASnotyetII}, the
second paper of the series, we will present the derived properties of
the UCM galaxies using these data and techniques. Paper~II will
discuss the young and newly-formed stars in the galaxies, together
with the underlying population of evolved stars, the total stellar
masses, etc.

This paper is structured as follows: section~\ref{data} introduces the
sample, the observations and the data measurements. Section
\ref{method} describes our modelling techniques, including the main
features of the stellar and nebular emission models, the star
formation scenarios and the extinction
prescriptions. Section~\ref{discussion} discusses the goodness of the
fits and possible correlations with the input data. Finally,
Section~\ref{summary} summarizes our conclusions. Throughout this
paper we use a cosmology with $\mathrm
H_{0}=70$\,km\,s$^{-1}$\,Mpc$^{-1}$, $\Omega_{\mathrm M}$=0.3 and
$\Lambda$=0.7.


\section{Data}
\label{data}

\subsection{The sample}

The UCM Survey galaxy sample contains 191 galaxies selected by their
$\mathrm H\alpha$ emission at an average redshift of 0.026
\citep{1994ApJS...95..387Z,1996ApJS..105..343Z,1996A&AS..120..323G}. Out of these
galaxies, 15 were classified as active galactic nuclei (AGN, including
Seyfert 1, Seyfert 2 and LINER types) by
\citet{1996A&AS..120..323G}, and will be excluded from the present study. The
rest are star-forming galaxies. Eleven of these were observed only in two
photometric bands, and   comparison with the models has not been attempted.
Thus, the sample studied here contains 163 galaxies, i.e., 94\% of all the
star-forming galaxies in the  complete UCM sample. This represents a factor of
2.5 increase over the sample studied by \gil.

The sample contains low excitation, high metallicity objects (often
with bright and dusty starbursts) and high excitation, low metallicity
ones with blue star-forming knots which may sometimes dominate the
optical luminosity of the whole galaxy, as in the case of {\it Blue
Compact Dwarfs}---BCDs. These two global spectroscopic types will be
called disk-{\it like} and HII-{\it like} galaxies,
respectively. There is also a large spectrum of sizes and masses (from
grand-design spirals to dwarfs), luminosities, emission-line
equivalent widths and star formation rates (SFRs).  The data required
in the present work are available for 94\% of the entire UCM sample
(excluding AGN). The galaxies studied here are thus a virtually
complete sample, with no biases against any of the previously
mentioned properties.

The dataset used in this work comprises a great deal of observations, both
photometric and spectroscopic. Most of them have already been presented in
previous papers. Only near infrared (nIR) data for the whole UCM Survey has not
been described before. In the next subsections we will review all these data,
with special emphasis on the nIR campaigns.

\subsection{Imaging}

\subsubsection{Optical: $B$- and $r$-bands}

Gunn $r$ and Johnson $B$ observations were obtained in several
observing runs from 1989 to 2001 using 1-2 metre-class telescopes at
the German-Spanish Observatory of Calar Alto (CAHA, Almer\'{\i}a,
Spain) and the Observatorio del Roque de los Muchachos (La Palma,
Spain). The observing details as well as the reduction and calibration
procedures can be found in 
\citet{1996A&AS..118....7V,1996A&AS..120..385V} for the $r$ data, and
\citet{2000A&AS..141..409P} and \citet{2001A&A...365..370P} for $B$.
Briefly, the sample has average magnitudes of $m_B$=16.1$\pm$1.1
($M_B=-19.2$) and $m_r=15.5\pm1.0$ ($M_r=-19.8$), with a mean $B$-band
effective radius of 2.8~kpc. Up to 65\% of the sample galaxies are
classified as Sb or later.

\subsubsection{Near infrared: $J$- and $K$-bands}

Near infrared observations for a small subsample of 67 galaxies were
presented in \gil. The whole sample of 191 galaxies has now been
observed in the nIR.

A total number of 11 campaigns were necessary to complete the 191
objects.  These runs were carried out from January 1996 to April 2002
in 1-2 metre-class telescopes: the 2.2m Telescope at Calar Alto
Observatory (Almer\'{\i}a, Spain), the 1m Telescope at UCO/Lick
Observatory (California,USA) and the 2.3m Bok Telescope of the
University of Arizona on Kitt Peak Observatory (Arizona, USA). Basic
information on each observing run are given in
Table~\ref{nirdata}. The filters used in these runs were $J$, $K$,
$K_s$ and $K'$. Standard reduction procedures in the nIR were applied,
a description of which can be found in
\citet{1993MNRAS.262..764A}. Flux calibration was performed using
standard stars from the lists of
\citet{1982AJ.....87.1029E,1998AJ....115.2594H,2001MNRAS.325..563H}. 
For each photometric night, appropriate atmospheric extinction
coefficients were derived and zero-points for each observing setup
determined. Non-photometric data were calibrated using short exposures
of the fields taken during photometric nights. The magnitudes of the
62 galaxies observed in $K'$ were transformed into the standard $K$
system by applying the constant offset $K'-K$=0.07 mag
\citep{1992AJ....103..332W,1993MNRAS.262..764A}. The correction from
$K_s$ to $K$ is negligible \citep{1998AJ....116.2475P}.

\begin{table*}
\caption{Log of the nIR observation for the UCM sample.}
\label{nirdata}
\begin{tabular}[c]{lcccl}
\hline
Telesc./Observ. & Dates & Chip &Plate scale&
\multicolumn{1}{c}{Conditions}\\ (1) & (2) & (3) & (4)
&\multicolumn{1}{c}{(5)}\\
\hline				                       			     
\hline
Lick 1.0m       & Jan  9-14      1996 &   NICMOS3 256x256 &    0.57   & 3 photometric nights\\
Lick 1.0m       & May  4-7       1996 &   NICMOS3 256x256 &    0.57   & 2 photometric nights\\
Lick 1.0m       & Jun  7-9       1996 &   NICMOS3 256x256 &    0.57   & photometric         \\
CAHA 2.2m       & Aug  4-6       1996 &   NICMOS3 256x256 &    0.63   & photometric         \\
Bok 2.3m        & Jan 10-17      1998 &   NICMOS3 256x256 &    0.60   & 2 photometric nights\\
Bok 2.3m        & Nov 01-07      1998 &   NICMOS3 256x256 &    0.60   & photometric         \\
Bok 2.3m        & Mar 20-23      1999 &   NICMOS3 256x256 &    0.60   & photometric         \\
Bok 2.3m        & Sep 27-30      1999 &   NICMOS3 256x256 &    0.60   & rainy               \\
Bok 2.3m        & Nov 07-09      2000 &   NICMOS3 256x256 &    0.60   & photometric         \\
Bok 2.3m        & Nov 29- Dec 01 2001 &   NICMOS3 256x256 &    0.60   & 1 photometric night \\
Bok 2.3m        & Mar 30- Apr 01 2002 &   NICMOS3 256x256 &    0.60   & photometric         \\
\hline	
\end{tabular}
\setcounter{table}{0}
\caption{Observing log for the nIR observations of the UCM Survey galaxies.
Columns stand for: (1) Telescope name. (2) Date of the
observation. (3) Detector used. (4) Scale of the chip in
arcsec$\cdot$pixel$^{-1}$. (5) Weather conditions.}
\end{table*}

\subsection{Long-slit optical spectroscopy}

We use redshifts, $\mathrm{H\alpha+[NII]}$ equivalent widths ($EW$),
$\mathrm{H\alpha/[NII]}$ and $\mathrm H\alpha/\mathrm H\beta$
intensity ratios, and spectroscopic types from
\citet{1996A&AS..120..323G}. The $EW(\mathrm{H\alpha+[NII]})$ was transformed into 
$EW(\mathrm H\alpha)$ using the observed $\mathrm{H\alpha/[NII]}$
ratios when available.  For the 20 galaxies without measured
$\mathrm{[NII]/H\alpha}$ ratios we assumed average values for the
relevant spectroscopic types.  Errors for the $\mathrm H\alpha$
equivalent width are estimated to be $\simeq20$\%.

For 30 objects, the $\mathrm H\alpha/H\beta$ ratio was impossible to
measure due to high extinction and/or to stellar absorption leading to
the absence of detectable $\mathrm H\beta$ emission. In these cases, the
average value of the 25\% highest ratios for each spectroscopic type
has been assumed. The rationale behind this assumption is that these
galaxies must have high extinctions in order to completely obliterate
the $\mathrm H\beta$ emission line.

The emission-line data was corrected for underlying stellar population
absorption. \citet{1992IAUS..149..225K} established that the $\mathrm
H\alpha$ and $\mathrm H\beta$ equivalent widths are equal within a 30\%
uncertainty. Thus, we used a typical stellar absorption equivalent
width for both $\mathrm H\alpha$ and $\mathrm H\beta$ of 3\AA\,
(\citealt{1998ApJS..116....1T,1999ApJS..125..489G}).


Although described elsewhere (see
\citealt{1996A&AS..120..323G} for details), we  
outline here the main properties of the different 
spectroscopic types that will be used
later in the discussion and in \pii:

	\noindent {\bf\large SBN} ---{\it Starburst Nuclei}---
	Originally defined by \citet{1983ApJ...268..602B}, they show
	high extinction values, with very low $\mathrm{[NII]/H\alpha}$
	ratios and faint $\mathrm{[OIII]\lambda5007}$ emission. Their
	$\mathrm H\alpha$ luminosities are always higher than
	$10^{8}\,L_{\sun}$.

        \noindent {\bf\large DANS} ---{\it Dwarf Amorphous Nuclear
        Starburst}--- Introduced by \citet{1989ApJS...70..479S}, they
        show very similar spectroscopic properties to SBN objects, but
        with $\mathrm H\alpha$ luminosities below
        $5\cdot10^{7}\,L_{\sun}$.

        \noindent {\bf\large HIIH} ---{\it HII Hotspot}--- The HII
        Hotspot class shows similar $\mathrm H\alpha$ luminosities to
        those measured in SBN galaxies but with large
        $\mathrm{[OIII]\lambda5007/H\beta}$ ratios, that is,
        higher ionization.

        \noindent {\bf\large DHIIH} ---{\it Dwarf HII Hotspot}----
        This is an HIIH subclass with identical spectroscopic
        properties but $\mathrm H\alpha$ luminosities lower than
        $5\cdot10^{7}\,L_{\sun}$.

        \noindent {\bf\large BCD} ---{\it Blue Compact Dwarf}--- The
        lowest luminosity and highest ionization objects have been
        classified as Blue Compact Dwarf galaxies. They show in all
        cases $\mathrm H\alpha$ luminosities lower than
        $5\cdot10^{7}\,L_{\sun}$ as well as large
        $\mathrm{[OIII]\lambda5007/H\beta}$ and
        $\mathrm{H\alpha/[NII]\lambda6584}$ line ratios and intense
        $\mathrm{[OII]\lambda3727}$ emission.

All these spectroscopic classes are usually collapsed in two main
categories: disk-{\it like} and HII-{\it like} galaxies (see
\citealt{1997ApJ...489..559G} and \citealt{1998Ap&SS.263....1G}).  
The disk-{\it like} class includes SBN and DANS spectroscopic types,
whereas the HII-{\it like} includes HIIH, DHIIH and BCD galaxies.

\subsection{Photometry analysis}
\label{photo}

Standard reduction procedures were applied to each photometric
dataset. The sky level was measured using $\sim30$ circular apertures
of $\sim100$ pixels$^2$ area placed at different positions around each
object. The average of all the measurements and its standard deviation
were used to determine the sky background and the related uncertainty.

In order to study the integrated properties of the galaxies, aperture
photometry was obtained for each bandpass. Aiming at including the
majority of the galaxy light, we used apertures with radii equal to
three exponential disk scale lengths as determined in the $r$-band
images (\citealt{1996A&AS..120..385V}). In the few cases when the
$r$-band bulge-disk decomposition was not available, we used the
radius of the 24 mag$\cdot$arcsec$^2$ isophote measured in the
$B$-band ($r_{24}$, \citealt{2001A&A...365..370P}). We inspected each
image visually and checked that these apertures were encompassing all
the detectable galaxy flux, and that no artifacts were disturbing the
data. In a few cases we slightly decreased or increased the aperture
radius in order to ensure that the measured flux was as close as
possible to the total flux. The photometric apertures were centred on
the peak of the galaxy light in each band. We checked that the effect
of possible misalignments between the light peaks in the different
bands was always below the photometric uncertainty. We estimate that
the size of this effect is always below 0.05 mag in $B$ and $r$ and
0.1 mag in $J$ and $K$.

Total $K$-band magnitudes were determined interactively as the average
of the measurements inside the outer apertures where the curve of
growth was flat. These fluxes were converted to absolute magnitudes
and corrected for Galactic extinction using the maps published by
\citet{1998ApJ...500..525S}.

Since the model-fitting procedure (explained in Sect.~\ref{fitting})
takes into account the observational errors, we took special care in
their determination.  The main sources of uncertainty are
photon-counting errors (described by Poisson statistics), readout
noise, flat-field errors (affecting mainly the sky determination), and
photometric calibration uncertainties. For a given aperture, the
uncertainty due to photon-counting errors and readout noise can be
written as:
\begin{equation}
\sigma_{\mathrm Poisson}=\frac{\sqrt{\left(C_{\mathrm gal}+n_{\mathrm gal}\cdot C_{\mathrm sky}\right)\cdot G+n_{\mathrm gal}\cdot RON^2}}{G}
\end{equation}
where $C_{\mathrm gal}$ is the number of counts coming from the galaxy,
$n_{\mathrm gal}$ the number of pixels inside the aperture, $C_{\mathrm sky}$
the sky value in counts, and $G$ and $RON$ the gain and readout noise
of the detector, measured in electrons/pixel and electrons,
respectively.

The error in the total flux arising from the determination of the sky
value is 
\begin{equation}
\sigma_{\mathrm sky det.}=\sigma_{\mathrm sky}\cdot n_{\mathrm gal}
\end{equation}
where $\sigma_{\mathrm sky}$ is the standard deviation of the sky 
measurements mentioned before.

Expressing the previous uncertainties in magnitudes, we get
\begin{equation}
\Delta m_{\mathrm image}=1.0857\cdot\frac{\sqrt{\left(\sigma_{\mathrm Poisson}^2+\sigma_{\mathrm sky det.}^2\right)}}{C_{\mathrm gal}\cdot \sqrt{N_{\mathrm im}}}
\end{equation}
where $N_{\mathrm im}$ is the number of images of the same object, ranging
from 1 in the optical filters to 20-24 in the nIR ones.

Finally, this quantity must be combined with the standard deviation of
the photometric calibration ($\sigma_{\mathrm zero-point}$) to obtain the
total error in the magnitudes:
\begin{equation}
\Delta m_{T}=\sqrt{(\Delta m_{\mathrm image})^2+\sigma_{\mathrm zero-point}^2}
\end{equation}
Typical total errors are 0.04 mag in $B$, 0.03 mag in $r$ and
0.09 mag in $J$ and $K$.

\begin{table*}\tabcolsep=0.1cm
\setcounter{table}{1} 
\caption{Photometric and spectroscopic data for the whole UCM sample.} 
\begin{tabular}{lcrrrrrrrrllr} 
\hline 
UCM name    &    z     &     \multicolumn{1}{c}{$m_B$}      &      \multicolumn{1}{c}{$m_r$}     &     \multicolumn{1}{c}{$m_J$}     &      \multicolumn{1}{c}{$m_K$}     &$EW(\mathrm H\alpha)$&$3d_L$ (kpc)&  $\frac{F_{\mathrm H\alpha}}{F_{\mathrm H\beta}}$ & $A_V^{\mathrm Gal}$ &  MphT  &  SpT   &  $M_K$\\ 
 \multicolumn{1}{c}{(1)} & \multicolumn{1}{c}{(2)} & \multicolumn{1}{c}{(3)} & \multicolumn{1}{c}{(4)} & \multicolumn{1}{c}{(5)} & \multicolumn{1}{c}{(6)} & \multicolumn{1}{c}{(7)} &\multicolumn{1}{c}{(8)} & (9) & (10) & (11) & (12) & (13) \\
\hline 
\hline
0000$+$2140   &  0.0238  &  14.61$\pm$0.03  &   \multicolumn{1}{c}{$-$}  &  11.71$\pm$0.11  &  10.37$\pm$0.03   &   103$\pm$21  &  13.9   &   7.55  &  0.15  & INTER & HIIH    &  $-$24.73 \\
0003$+$2200   &  0.0224  &  17.19$\pm$0.02  &  16.30$\pm$0.04  &  14.65$\pm$0.15  &  13.53$\pm$0.08   &    38$\pm$ 8  &   9.9   &   5.02  &  0.23  & Sc$+$   & DANS    &  $-$21.47 \\
0003$+$2215   &  0.0223  &  15.89$\pm$0.02  &   \multicolumn{1}{c}{$-$}  &   \multicolumn{1}{c}{$-$}  &  11.36$\pm$0.03   &    25$\pm$ 5  &   7.6   &   5.62  &  0.24  & Sc$+$   & SBN     &  $-$23.62 \\
0003$+$1955   &  0.0278  &  14.11$\pm$0.13  &   \multicolumn{1}{c}{$-$}  &   \multicolumn{1}{c}{$-$}  &   \multicolumn{1}{c}{$-$}   &   294$\pm$59  &   6.7   &   4.61  &  0.12  & $-$     & Sy1     &   \multicolumn{1}{c}{$-$}\\
0005$+$1802   &  0.0187  &  16.40$\pm$0.13  &   \multicolumn{1}{c}{$-$}  &   \multicolumn{1}{c}{$-$}  &  12.27$\pm$0.04   &    13$\pm$ 3  &   6.4   &   3.62  &  0.12  & Sb    & SBN     &  $-$22.31 \\
0006$+$2332   &  0.0159  &  14.95$\pm$0.05  &   \multicolumn{1}{c}{$-$}  &  12.69$\pm$0.05  &  11.90$\pm$0.06   &    58$\pm$12  &   8.4   &   4.58  &  0.31  & Sb    & HIIH    &  $-$22.37 \\
0013$+$1942   &  0.0272  &  17.13$\pm$0.02  &  16.60$\pm$0.03  &  15.03$\pm$0.07  &  14.07$\pm$0.06   &   124$\pm$25  &  13.2   &   3.61  &  0.13  & Sc$+$   & HIIH    &  $-$21.34 \\
0014$+$1829   &  0.0182  &  16.50$\pm$0.03  &  15.91$\pm$0.04  &  14.66$\pm$0.15  &  13.65$\pm$0.18   &   131$\pm$26  &   5.7   &   9.80  &  0.16  & Sa    & HIIH    &  $-$21.53 \\
0014$+$1748   &  0.0182  &  14.83$\pm$0.05  &  14.01$\pm$0.14  &  11.86$\pm$0.11  &  10.82$\pm$0.16   &    86$\pm$17  &  39.7   &   5.74  &  0.13  & SBb   & SBN     &  $-$23.71 \\
0015$+$2212   &  0.0198  &  16.85$\pm$0.02  &  16.04$\pm$0.08  &  14.30$\pm$0.07  &  13.29$\pm$0.04   &   120$\pm$24  &   8.5   &   3.32  &  0.23  & Sa    & HIIH    &  $-$21.56 \\
0017$+$1942   &  0.0260  &  15.91$\pm$0.02  &  15.38$\pm$0.07  &  14.01$\pm$0.08  &  13.09$\pm$0.07   &   100$\pm$20  &  18.7   &   4.37  &  0.17  & Sc$+$   & HIIH    &  $-$22.29 \\
0017$+$2148   &  0.0189  &  16.95$\pm$0.05  &   \multicolumn{1}{c}{$-$}  &  14.31$\pm$0.24  &  13.30$\pm$0.04   &    74$\pm$15  &   3.0   &   4.66  &  0.21  & Sa    & HIIH    &  $-$21.43 \\
0018$+$2216   &  0.0169  &  16.95$\pm$0.02  &  16.15$\pm$0.03  &  14.22$\pm$0.07  &  13.39$\pm$0.05   &    15$\pm$ 3  &   5.7   &   2.86  &  0.23  & Sb    & DANS    &  $-$21.08 \\
0018$+$2218   &  0.0220  &  15.97$\pm$0.02  &   \multicolumn{1}{c}{$-$}  &  12.17$\pm$0.14  &  11.12$\pm$0.20   &    16$\pm$ 3  &  10.8   &   9.39  &  0.22  & Sb    & SBN     &  $-$23.81 \\
0019$+$2201   &  0.0191  &  16.80$\pm$0.02  &  15.82$\pm$0.04  &  13.96$\pm$0.04  &  12.96$\pm$0.05   &    33$\pm$ 7  &  10.4   &   3.70  &  0.21  & Sb    & DANS    &  $-$21.69 \\
0022$+$2049   &  0.0185  &  15.86$\pm$0.05  &  14.65$\pm$0.03  &  12.46$\pm$0.08  &  11.24$\pm$0.05   &    76$\pm$15  &  10.2   &   6.28  &  0.30  & Sb    & HIIH    &  $-$23.42 \\
0023$+$1908   &  0.0251  &  16.83$\pm$0.05  &   \multicolumn{1}{c}{$-$}  &  14.66$\pm$0.31  &  13.83$\pm$0.07   &   121$\pm$24  &   3.2   &   4.08  &  0.19  & Sc$+$   & HIIH    &  $-$21.39 \\
0034$+$2119   &  0.0315  &  15.86$\pm$0.03  &   \multicolumn{1}{c}{$-$}  &   \multicolumn{1}{c}{$-$}  &  11.84$\pm$0.07   &    19$\pm$ 4  &  12.2   &   3.58  &  0.11  & SBc$+$  & SBN     &  $-$23.91 \\
0037$+$2226   &  0.0195  &  14.65$\pm$0.05  &   \multicolumn{1}{c}{$-$}  &  12.44$\pm$0.13  &  11.53$\pm$0.03   &    45$\pm$ 9  &   7.7   &   4.19  &  0.13  & SBc$+$  & SBN     &  $-$23.23 \\
0038$+$2259   &  0.0464  &  16.39$\pm$0.05  &  15.61$\pm$0.04  &  13.84$\pm$0.26  &  12.99$\pm$0.04   &    21$\pm$ 4  &  33.8   &   4.63  &  0.09  & Sb    & SBN     &  $-$23.60 \\
0039$+$0054   &  0.0191  &  15.22$\pm$0.05  &   \multicolumn{1}{c}{$-$}  &   \multicolumn{1}{c}{$-$}  &  11.93$\pm$0.07   &    23$\pm$ 5  &   8.8   &   8.75  &  0.07  & Sc$+$   & SBN     &  $-$22.74 \\
0040$+$0257   &  0.0367  &  16.98$\pm$0.05  &  16.85$\pm$0.04  &   \multicolumn{1}{c}{$-$}  &  14.41$\pm$0.08   &   119$\pm$24  &  12.5   &   4.14  &  0.09  & Sb    & DANS    &  $-$21.64 \\
0040$+$2312   &  0.0254  &  15.69$\pm$0.03  &   \multicolumn{1}{c}{$-$}  &  12.15$\pm$0.14  &  11.07$\pm$0.03   &    28$\pm$ 6  &  12.9   &   8.55  &  0.12  & Sc$+$   & SBN     &  $-$24.22 \\
0040$+$0220   &  0.0173  &  17.25$\pm$0.15  &  16.61$\pm$0.03  &  15.16$\pm$0.04  &  14.23$\pm$0.03   &    77$\pm$15  &   4.4   &   3.86  &  0.07  & Sc$+$   & DANS    &  $-$20.23 \\
0040$-$0023   &  0.0142  &  13.76$\pm$0.03  &   \multicolumn{1}{c}{$-$}  &  11.15$\pm$0.10  &  10.35$\pm$0.07   &    18$\pm$ 4  &  10.8   &   9.20  &  0.06  & Sc$+$   & LINER   &  $-$23.60 \\
0041$+$0134   &  0.0169  &  14.42$\pm$0.04  &   \multicolumn{1}{c}{$-$}  &   \multicolumn{1}{c}{$-$}  &  11.46$\pm$0.08   &    12$\pm$ 2  &  13.3   &   8.96  &  0.08  & Sc$+$   & SBN     &  $-$22.87 \\
0043$+$0245   &  0.0180  &  17.34$\pm$0.05  &   \multicolumn{1}{c}{$-$}  &   \multicolumn{1}{c}{$-$}  &  14.30$\pm$0.08   &    34$\pm$ 7  &   2.2   &   5.07  &  0.07  & Sc$+$   & HIIH    &  $-$20.26 \\
0043$-$0159   &  0.0161  &  13.01$\pm$0.05  &   \multicolumn{1}{c}{$-$}  &  10.79$\pm$0.01  &   9.70$\pm$0.07   &    60$\pm$12  &   9.8   &   8.03  &  0.09  & Sc$+$   & SBN     &  $-$24.53 \\
0044$+$2246   &  0.0253  &  16.06$\pm$0.15  &  14.90$\pm$0.08  &  12.54$\pm$0.07  &  11.47$\pm$0.05   &    25$\pm$ 5  &  33.8   &   7.42  &  0.12  & Sb    & SBN     &  $-$23.78 \\
0045$+$2206   &  0.0203  &  15.06$\pm$0.05  &   \multicolumn{1}{c}{$-$}  &  12.94$\pm$0.07  &  12.04$\pm$0.05   &    80$\pm$16  &   5.6   &   4.14  &  0.15  & INTER & HIIH    &  $-$22.71 \\
0047$+$2051   &  0.0577  &  16.98$\pm$0.05  &  16.14$\pm$0.03  &   \multicolumn{1}{c}{$-$}  &  13.13$\pm$0.03   &    73$\pm$15  &  20.0   &   4.60  &  0.10  & Sc$+$   & SBN     &  $-$23.96 \\
0047$-$0213   &  0.0144  &  15.73$\pm$0.04  &  14.97$\pm$0.04  &  13.13$\pm$0.13  &  12.25$\pm$0.04   &    40$\pm$ 8  &  10.5   &   4.94  &  0.15  & S0    & DHIIH   &  $-$21.94 \\
0047$+$2413   &  0.0347  &  15.88$\pm$0.05  &  14.81$\pm$0.03  &  12.74$\pm$0.05  &  11.63$\pm$0.05   &    61$\pm$12  &  31.4   &   5.13  &  0.20  & Sa    & SBN     &  $-$24.39 \\
0047$+$2414   &  0.0347  &  15.22$\pm$0.05  &   \multicolumn{1}{c}{$-$}  &  12.66$\pm$0.18  &  11.69$\pm$0.03   &    78$\pm$16  &  10.1   &   4.69  &  0.20  & Sc$+$   & SBN     &  $-$24.28 \\
0049$-$0006   &  0.0377  &  18.68$\pm$0.05  &  18.52$\pm$0.04  &  17.80$\pm$0.09  &  16.62$\pm$0.14   &   346$\pm$69  &   7.4   &   2.86  &  0.08  & BCD   & BCD     &  $-$19.50 \\
0049$+$0017   &  0.0140  &  17.19$\pm$0.03  &  16.69$\pm$0.09  &  15.36$\pm$0.05  &  14.50$\pm$0.07   &   310$\pm$62  &   6.2   &   2.86  &  0.08  & Sb    & DHIIH   &  $-$19.42 \\
0049$-$0045   &  0.0055  &  15.34$\pm$0.02  &   \multicolumn{1}{c}{$-$}  &  13.05$\pm$0.15  &  12.31$\pm$0.07   &    73$\pm$15  &   1.6   &   4.79  &  0.13  & Sb    & HIIH    &  $-$19.73 \\
0050$+$0005   &  0.0346  &  16.54$\pm$0.03  &  16.03$\pm$0.03  &   \multicolumn{1}{c}{$-$}  &  13.68$\pm$0.07   &    94$\pm$19  &  13.1   &   4.50  &  0.08  & Sa    & HIIH    &  $-$22.31 \\
0050$+$2114   &  0.0245  &  15.56$\pm$0.05  &  14.78$\pm$0.03  &  12.76$\pm$0.09  &  11.59$\pm$0.09   &    69$\pm$14  &  15.5   &   5.73  &  0.13  & Sa    & SBN     &  $-$23.59 \\
0051$+$2430   &  0.0173  &  15.40$\pm$0.15  &   \multicolumn{1}{c}{$-$}  &  11.94$\pm$0.09  &  11.06$\pm$0.04   &    34$\pm$ 7  &   5.7   &   6.12  &  0.15  & Sa    & SBN     &  $-$23.34 \\
0054$-$0133   &  0.0512  &  16.00$\pm$0.04  &   \multicolumn{1}{c}{$-$}  &  12.99$\pm$0.13  &  11.80$\pm$0.07   &    23$\pm$ 4  &  13.4   &   8.79  &  0.12  & Sb    & SBN     &  $-$25.02 \\
0054$+$2337   &  0.0164  &  15.27$\pm$0.03  &   \multicolumn{1}{c}{$-$}  &  13.27$\pm$0.09  &  12.66$\pm$0.09   &    62$\pm$12  &   6.2   &   4.68  &  0.16  & Sc$+$   & HIIH    &  $-$21.67 \\
0056$+$0044   &  0.0183  &  16.82$\pm$0.05  &  16.52$\pm$0.10  &  15.55$\pm$0.15  &  14.54$\pm$0.16   &   399$\pm$80  &  17.7   &   3.03  &  0.09  & Irr   & DHIIH   &  $-$20.04 \\
0056$+$0043   &  0.0189  &  16.63$\pm$0.05  &  16.20$\pm$0.03  &   \multicolumn{1}{c}{$-$}  &  13.88$\pm$0.07   &    53$\pm$11  &   6.8   &   3.81  &  0.09  & Sb    & DHIIH   &  $-$20.77 \\
0119$+$2156   &  0.0583  &  16.66$\pm$0.29  &  15.46$\pm$0.10  &  13.31$\pm$0.05  &  11.93$\pm$0.04   &    16$\pm$ 3  &  145.6   &   7.89  &  0.17  & Sb    & Sy2     &  $-$25.20 \\
0121$+$2137   &  0.0345  &  16.02$\pm$0.05  &  15.47$\pm$0.06  &  13.85$\pm$0.08  &  12.90$\pm$0.07   &    66$\pm$13  &  33.8   &   4.86  &  0.22  & Sc$+$   & SBN     &  $-$23.05 \\
0129$+$2109   &  0.0344  &  15.01$\pm$0.04  &   \multicolumn{1}{c}{$-$}  &  12.06$\pm$0.07  &  11.00$\pm$0.05   &    32$\pm$ 6  &  14.4   &   8.41  &  0.19  & SBc$+$  & LINER   &  $-$24.95 \\
0134$+$2257   &  0.0353  &  16.03$\pm$0.05  &   \multicolumn{1}{c}{$-$}  &  12.76$\pm$0.13  &  11.73$\pm$0.03   &    26$\pm$ 5  &  10.6   &   4.91  &  0.37  & Sb    & SBN     &  $-$24.40 \\
0135$+$2242   &  0.0363  &  17.16$\pm$0.05  &  16.26$\pm$0.03  &  14.40$\pm$0.04  &  13.42$\pm$0.05   &    46$\pm$ 9  &  14.4   &   6.69  &  0.40  & S0    & DANS    &  $-$22.74 \\
0138$+$2216   &  0.0591  &  17.71$\pm$0.03  &   \multicolumn{1}{c}{$-$}  &  14.35$\pm$0.20  &  13.18$\pm$0.07   &    10$\pm$ 2  &   7.4   &   3.35  &  0.39  & Sc$+$   & $-$       &  $-$24.11 \\
0141$+$2220   &  0.0174  &  16.36$\pm$0.05  &  15.91$\pm$0.03  &  13.72$\pm$0.04  &  12.66$\pm$0.02   &    37$\pm$ 7  &   9.0   &   4.68  &  0.30  & Sa    & DANS    &  $-$21.88 \\
0142$+$2137   &  0.0362  &  15.35$\pm$0.05  &  14.25$\pm$0.05  &   \multicolumn{1}{c}{$-$}  &  11.19$\pm$0.04   &    29$\pm$ 6  &  48.3   &   3.83  &  0.34  & SBb   & Sy2     &  $-$24.98 \\
0144$+$2519   &  0.0409  &  15.67$\pm$0.05  &  14.98$\pm$0.06  &  13.12$\pm$0.11  &  12.13$\pm$0.12   &    29$\pm$ 6  &  38.2   &   5.66  &  0.42  & SBc$+$  & SBN     &  $-$24.20 \\
0147$+$2309   &  0.0194  &  16.88$\pm$0.05  &  15.99$\pm$0.04  &  14.56$\pm$0.05  &  13.62$\pm$0.06   &   118$\pm$24  &  10.8   &   4.34  &  0.32  & Sa    & HIIH    &  $-$21.05 \\
0148$+$2124   &  0.0169  &  17.19$\pm$0.05  &  16.49$\pm$0.03  &  15.23$\pm$0.04  &  14.43$\pm$0.06   &   136$\pm$27  &   6.2   &   3.26  &  0.21  & BCD   & BCD     &  $-$20.00 \\
0150$+$2032   &  0.0323  &  16.46$\pm$0.15  &  16.19$\pm$0.10  &  15.07$\pm$0.40  &  13.49$\pm$0.08   &   171$\pm$34  &  29.9   &   3.34  &  0.25  & Sc$+$   & HIIH    &  $-$22.42 \\
0156$+$2410   &  0.0134  &  15.33$\pm$0.04  &  14.66$\pm$0.03  &  13.02$\pm$0.04  &  12.24$\pm$0.05   &    40$\pm$ 8  &  10.9   &   4.45  &  0.31  & Sb    & DANS    &  $-$21.70 \\
0157$+$2413   &  0.0177  &  15.08$\pm$0.09  &  13.79$\pm$0.04  &  11.08$\pm$0.07  &  10.36$\pm$0.03   &    25$\pm$ 5  &  30.1   &   5.03  &  0.33  & Sc$+$   & Sy2     &  $-$24.16 \\
0157$+$2102   &  0.0106  &  15.01$\pm$0.04  &  14.58$\pm$0.03  &  13.01$\pm$0.04  &  12.31$\pm$0.05   &    61$\pm$12  &   7.6   &   3.89  &  0.29  & Sb    & HIIH    &  $-$21.10 \\
0159$+$2354   &  0.0170  &  17.34$\pm$0.05  &  16.36$\pm$0.03  &  14.50$\pm$0.04  &  13.59$\pm$0.05   &    63$\pm$13  &   6.6   &   4.18  &  0.33  & Sb    & HIIH    &  $-$20.86 \\
0159$+$2326   &  0.0178  &  16.01$\pm$0.05  &  14.87$\pm$0.03  &  12.78$\pm$0.07  &  11.84$\pm$0.05   &    28$\pm$ 6  &  12.1   &   6.14  &  0.28  & Sc$+$   & DANS    &  $-$22.82 \\
1246$+$2727   &  0.0199  &  15.84$\pm$0.21  &   \multicolumn{1}{c}{$-$}  &  13.82$\pm$0.35  &  12.92$\pm$0.09   &    67$\pm$13  &   6.7   &   4.90  &  0.04  & Irr   & HIIH    &  $-$21.85 \\
1247$+$2701   &  0.0231  &  16.76$\pm$0.09  &  16.12$\pm$0.03  &  14.49$\pm$0.03  &  13.69$\pm$0.05   &    28$\pm$ 6  &  12.8   &   3.21  &  0.04  & Sc$+$   & DANS    &  $-$21.33 \\
\hline
\hline
\end{tabular}\\
\end{table*}\tabcolsep=0.1cm
\setcounter{table}{1}\begin{table*}
\caption{continued}
\begin{tabular}{lcrrrrrrrrllr} 
\hline 
UCM name    &    z     &     \multicolumn{1}{c}{$m_B$}      &      \multicolumn{1}{c}{$m_r$}     &     \multicolumn{1}{c}{$m_J$}     &      \multicolumn{1}{c}{$m_K$}     &$EW(\mathrm H\alpha)$&$3d_L$ (kpc)&  $\frac{F_{\mathrm H\alpha}}{F_{\mathrm H\beta}}$ & $A_V^{\mathrm Gal}$ &  MphT  &  SpT   &  $M_K$\\ 
 \multicolumn{1}{c}{(1)} & \multicolumn{1}{c}{(2)} & \multicolumn{1}{c}{(3)} & \multicolumn{1}{c}{(4)} & \multicolumn{1}{c}{(5)} & \multicolumn{1}{c}{(6)} & \multicolumn{1}{c}{(7)} &\multicolumn{1}{c}{(8)} & (9) & (10) & (11) & (12) & (13) \\
\hline 
\hline
1248$+$2912   &  0.0217  &  15.09$\pm$0.17  &   \multicolumn{1}{c}{$-$}  &   \multicolumn{1}{c}{$-$}  &  11.55$\pm$0.07   &    29$\pm$ 6  &   8.0   &   3.99  &  0.06  & SBb   & SBN     &  $-$23.33 \\
1253$+$2756   &  0.0165  &  16.09$\pm$0.02  &  15.41$\pm$0.03  &  13.99$\pm$0.05  &  13.12$\pm$0.04   &   114$\pm$23  &   6.0   &   2.86  &  0.03  & Sa    & HIIH    &  $-$21.59 \\
1254$+$2740   &  0.0161  &  16.25$\pm$0.03  &  15.54$\pm$0.04  &   \multicolumn{1}{c}{$-$}  &   \multicolumn{1}{c}{$-$}   &    58$\pm$12  &  18.5   &   4.28  &  0.04  & Sa    & SBN     &   \multicolumn{1}{c}{$-$}\\
1254$+$2802   &  0.0253  &  16.91$\pm$0.02  &  15.88$\pm$0.03  &  13.91$\pm$0.03  &  12.84$\pm$0.04   &    14$\pm$ 3  &  14.7   &   8.78  &  0.04  & Sc$+$   & DANS    &  $-$22.44 \\
1255$+$2819   &  0.0273  &  16.10$\pm$0.12  &  15.33$\pm$0.03  &  13.63$\pm$0.05  &  12.66$\pm$0.05   &    47$\pm$ 9  &  19.7   &   4.16  &  0.04  & Sb    & SBN     &  $-$23.10 \\
1255$+$3125   &  0.0258  &  16.46$\pm$0.13  &  15.30$\pm$0.03  &  13.44$\pm$0.14  &  12.55$\pm$0.17   &    64$\pm$13  &  12.6   &   3.92  &  0.06  & Sa    & HIIH    &  $-$22.77 \\
1255$+$2734   &  0.0234  &  16.97$\pm$0.02  &  16.15$\pm$0.03  &   \multicolumn{1}{c}{$-$}  &  13.33$\pm$0.06   &    99$\pm$20  &  10.6   &   5.44  &  0.04  & Sc$+$   & SBN     &  $-$21.74 \\
1256$+$2717   &  0.0273  &  17.93$\pm$0.13  &   \multicolumn{1}{c}{$-$}  &   \multicolumn{1}{c}{$-$}  &  15.35$\pm$0.14   &    62$\pm$12  &   3.6   &   3.85  &  0.03  & S0    & DHIIH   &  $-$20.04 \\
1256$+$2732   &  0.0245  &  15.95$\pm$0.18  &  15.37$\pm$0.04  &  13.90$\pm$0.05  &  12.90$\pm$0.07   &    79$\pm$16  &  31.0   &   4.71  &  0.05  & INTER & SBN     &  $-$22.26 \\
1256$+$2701   &  0.0247  &  16.66$\pm$0.09  &  16.27$\pm$0.07  &  14.70$\pm$0.10  &  13.68$\pm$0.11   &   109$\pm$22  &  32.5   &   3.46  &  0.03  & Sc$+$   & HIIH    &  $-$21.49 \\
1256$+$2910   &  0.0279  &  16.21$\pm$0.08  &  15.28$\pm$0.03  &  13.45$\pm$0.03  &  12.52$\pm$0.04   &    25$\pm$ 5  &  19.5   &   8.66  &  0.03  & Sb    & SBN     &  $-$23.16 \\
1256$+$2823   &  0.0315  &  16.14$\pm$0.10  &  15.30$\pm$0.03  &  13.67$\pm$0.10  &  12.50$\pm$0.14   &    76$\pm$15  &  16.9   &   4.82  &  0.04  & Sb    & SBN     &  $-$23.35 \\
1256$+$2754   &  0.0172  &  15.43$\pm$0.07  &  14.90$\pm$0.03  &  13.18$\pm$0.05  &  12.25$\pm$0.05   &    49$\pm$10  &  14.5   &   4.12  &  0.04  & Sa    & SBN     &  $-$22.44 \\
1256$+$2722   &  0.0287  &  17.21$\pm$0.09  &  16.21$\pm$0.04  &   \multicolumn{1}{c}{$-$}  &  12.84$\pm$0.06   &    26$\pm$ 5  &  14.3   &   5.10  &  0.04  & Sc$+$   & DANS    &  $-$22.66 \\
1257$+$2808   &  0.0171  &  16.38$\pm$0.02  &  15.66$\pm$0.03  &  14.26$\pm$0.32  &  12.91$\pm$0.29   &    29$\pm$ 6  &   7.2   &   5.57  &  0.03  & Sb    & SBN     &  $-$21.48 \\
1258$+$2754   &  0.0253  &  16.02$\pm$0.09  &  15.58$\pm$0.07  &   \multicolumn{1}{c}{$-$}  &  13.22$\pm$0.08   &   101$\pm$20  &  17.5   &   6.01  &  0.03  & Sb    & SBN     &  $-$22.06 \\
1259$+$2934   &  0.0239  &  13.99$\pm$0.09  &  12.85$\pm$0.03  &  10.78$\pm$0.05  &   9.78$\pm$0.04   &   148$\pm$30  &  43.3   &   7.75  &  0.04  & Sb    & Sy2     &  $-$25.37 \\
1259$+$3011   &  0.0307  &  16.25$\pm$0.09  &  15.40$\pm$0.03  &  13.56$\pm$0.13  &  12.57$\pm$0.14   &    22$\pm$ 4  &  36.5   &   3.50  &  0.04  & Sa    & SBN     &  $-$23.08 \\
1259$+$2755   &  0.0240  &  15.57$\pm$0.04  &  14.61$\pm$0.03  &  13.08$\pm$0.12  &  11.97$\pm$0.13   &    44$\pm$ 9  &  17.2   &   5.22  &  0.03  & Sa    & SBN     &  $-$23.25 \\
1300$+$2907   &  0.0219  &  17.27$\pm$0.09  &  16.86$\pm$0.03  &   \multicolumn{1}{c}{$-$}  &  14.75$\pm$0.10   &    94$\pm$19  &   9.7   &   5.10  &  0.04  & Sa    & HIIH    &  $-$20.16 \\
1301$+$2904   &  0.0266  &  15.97$\pm$0.10  &  15.57$\pm$0.03  &  14.07$\pm$0.05  &  13.39$\pm$0.06   &    69$\pm$14  &  16.1   &   3.13  &  0.04  & Sb    & HIIH    &  $-$22.03 \\
1302$+$2853   &  0.0237  &  16.50$\pm$0.02  &  15.99$\pm$0.03  &  14.26$\pm$0.14  &  13.43$\pm$0.19   &    40$\pm$ 8  &  10.1   &   4.07  &  0.04  & Sb    & DHIIH   &  $-$22.24 \\
1302$+$3032   &  0.0342  &  16.66$\pm$0.07  &   \multicolumn{1}{c}{$-$}  &  14.85$\pm$0.45  &  13.95$\pm$0.07   &    49$\pm$10  &   6.2   &   4.09  &  0.04  & Sa    & HIIH    &  $-$21.97 \\
1303$+$2908   &  0.0261  &  16.82$\pm$0.10  &  16.28$\pm$0.03  &  15.27$\pm$0.06  &  14.31$\pm$0.08   &   165$\pm$33  &  17.5   &   2.86  &  0.04  & Irr   & HIIH    &  $-$20.99 \\
1304$+$2808   &  0.0210  &  16.02$\pm$0.11  &  15.03$\pm$0.03  &  13.37$\pm$0.13  &  12.03$\pm$0.14   &    24$\pm$ 5  &  18.9   &   2.86  &  0.04  & Sb    & SBN     &  $-$22.83 \\
1304$+$2830   &  0.0217  &  18.62$\pm$0.04  &  18.09$\pm$0.03  &   \multicolumn{1}{c}{$-$}  &  15.43$\pm$0.09   &    56$\pm$11  &   4.7   &   3.57  &  0.04  & BCD   & DHIIH   &  $-$19.45 \\
1304$+$2907   &  0.0159  &  15.16$\pm$0.24  &  14.61$\pm$0.08  &   \multicolumn{1}{c}{$-$}  &  12.55$\pm$0.10   &     8$\pm$ 2  &  28.6   &   8.96  &  0.04  & Irr   & $-$       &  $-$21.64 \\
1304$+$2818   &  0.0244  &  15.88$\pm$0.02  &  15.06$\pm$0.03  &  13.58$\pm$0.06  &  12.50$\pm$0.08   &    80$\pm$16  &  18.5   &   2.97  &  0.05  & Sc$+$   & SBN     &  $-$22.72 \\
1306$+$2938   &  0.0209  &  15.59$\pm$0.03  &  15.09$\pm$0.03  &  13.60$\pm$0.05  &  12.37$\pm$0.06   &   100$\pm$20  &  10.6   &   3.93  &  0.04  & SBb   & SBN     &  $-$22.73 \\
1306$+$3111   &  0.0168  &  16.44$\pm$0.02  &  15.54$\pm$0.03  &  13.85$\pm$0.08  &  13.11$\pm$0.07   &    61$\pm$12  &   7.1   &   6.52  &  0.04  & Sc$+$   & DANS    &  $-$21.26 \\
1307$+$2910   &  0.0187  &  14.25$\pm$0.03  &  13.22$\pm$0.05  &  11.59$\pm$0.35  &  10.33$\pm$0.29   &    25$\pm$ 5  &  37.7   &   4.70  &  0.03  & SBb   & SBN     &  $-$24.22 \\
1308$+$2958   &  0.0212  &  15.36$\pm$0.02  &  14.53$\pm$0.04  &  12.71$\pm$0.08  &  11.94$\pm$0.15   &    21$\pm$ 4  &  27.1   &   5.63  &  0.04  & Sc$+$   & SBN     &  $-$22.89 \\
1308$+$2950   &  0.0242  &  14.91$\pm$0.13  &  13.90$\pm$0.04  &  11.83$\pm$0.11  &  10.77$\pm$0.18   &    37$\pm$ 7  &  49.3   &   8.84  &  0.04  & SBb   & SBN     &  $-$24.36 \\
1310$+$3027   &  0.0234  &  16.70$\pm$0.09  &  15.80$\pm$0.03  &  13.74$\pm$0.07  &  12.86$\pm$0.05   &    70$\pm$14  &  14.6   &   7.27  &  0.04  & Sb    & DANS    &  $-$22.33 \\
1312$+$3040   &  0.0233  &  15.71$\pm$0.09  &  14.80$\pm$0.03  &  12.94$\pm$0.05  &  11.74$\pm$0.07   &    53$\pm$11  &  16.6   &   3.82  &  0.04  & Sa    & SBN     &  $-$23.36 \\
1312$+$2954   &  0.0230  &  16.20$\pm$0.09  &  15.24$\pm$0.03  &  13.27$\pm$0.14  &  12.44$\pm$0.34   &    44$\pm$ 9  &  19.4   &   7.07  &  0.04  & Sc$+$   & SBN     &  $-$22.82 \\
1313$+$2938   &  0.0380  &  16.93$\pm$0.09  &  16.56$\pm$0.03  &  15.45$\pm$0.06  &  14.67$\pm$0.07   &   311$\pm$62  &   8.9   &   2.86  &  0.03  & Sa    & HIIH    &  $-$21.74 \\
1314$+$2827   &  0.0253  &  16.39$\pm$0.03  &  15.72$\pm$0.04  &   \multicolumn{1}{c}{$-$}  &  13.12$\pm$0.06   &    48$\pm$10  &  10.1   &   4.62  &  0.04  & Sa    & SBN     &  $-$22.30 \\
1320$+$2727   &  0.0247  &  17.51$\pm$0.13  &  17.08$\pm$0.03  &   \multicolumn{1}{c}{$-$}  &  14.86$\pm$0.08   &    52$\pm$10  &   7.9   &   2.98  &  0.06  & Sb    & DHIIH   &  $-$20.39 \\
1324$+$2926   &  0.0172  &  18.09$\pm$0.13  &  17.24$\pm$0.03  &  15.92$\pm$0.03  &  15.07$\pm$0.05   &   236$\pm$47  &   3.5   &   2.86  &  0.04  & BCD   & BCD     &  $-$19.49 \\
1324$+$2651   &  0.0249  &  15.20$\pm$0.13  &  14.56$\pm$0.03  &  13.01$\pm$0.03  &  11.89$\pm$0.04   &    75$\pm$15  &  19.0   &   4.74  &  0.04  & INTER & SBN     &  $-$23.37 \\
1331$+$2900   &  0.0356  &  19.11$\pm$0.13  &  18.62$\pm$0.03  &   \multicolumn{1}{c}{$-$}  &  17.29$\pm$0.26   &   549$\pm$110  &   5.9   &   2.86  &  0.04  & BCD   & BCD     &  $-$18.70 \\
1428$+$2727   &  0.0149  &  15.03$\pm$0.02  &  14.56$\pm$0.03  &  13.73$\pm$0.12  &  12.83$\pm$0.14   &   182$\pm$36  &   9.6   &   3.18  &  0.05  & Irr   & HIIH    &  $-$21.59 \\
1429$+$2645   &  0.0328  &  17.89$\pm$0.03  &  17.12$\pm$0.03  &  15.61$\pm$0.06  &  14.70$\pm$0.07   &    87$\pm$17  &  10.3   &   2.89  &  0.06  & Sb    & DHIIH   &  $-$21.24 \\
1430$+$2947   &  0.0290  &  16.53$\pm$0.11  &  15.92$\pm$0.03  &  14.47$\pm$0.06  &  13.57$\pm$0.09   &   132$\pm$26  &  20.9   &   3.69  &  0.06  & S0    & HIIH    &  $-$22.01 \\
1431$+$2854   &  0.0310  &  15.76$\pm$0.05  &  14.98$\pm$0.03  &  13.36$\pm$0.06  &  12.45$\pm$0.06   &    26$\pm$ 5  &  15.3   &   8.60  &  0.06  & Sb    & SBN     &  $-$23.34 \\
1431$+$2702   &  0.0384  &  17.31$\pm$0.02  &  16.76$\pm$0.03  &  15.10$\pm$0.08  &  14.13$\pm$0.04   &   134$\pm$27  &   8.6   &   3.50  &  0.06  & Sa    & HIIH    &  $-$22.18 \\
1431$+$2947   &  0.0219  &  17.92$\pm$0.06  &  17.53$\pm$0.03  &   \multicolumn{1}{c}{$-$}  &  15.76$\pm$0.17   &   131$\pm$26  &   9.7   &   2.86  &  0.05  & BCD   & BCD     &  $-$19.16 \\
1431$+$2814   &  0.0320  &  17.02$\pm$0.05  &  15.95$\pm$0.03  &  13.84$\pm$0.04  &  12.87$\pm$0.07   &    19$\pm$ 4  &  16.0   &   8.29  &  0.07  & Sb    & DANS    &  $-$22.91 \\
1432$+$2645   &  0.0307  &  15.40$\pm$0.03  &  14.60$\pm$0.03  &  12.88$\pm$0.13  &  11.78$\pm$0.18   &    34$\pm$ 7  &  42.2   &   4.88  &  0.09  & SBb   & SBN     &  $-$23.87 \\
1440$+$2521N  &  0.0315  &  16.85$\pm$0.02  &  15.85$\pm$0.03  &  13.69$\pm$0.32  &  12.63$\pm$0.28   &    54$\pm$11  &  16.7   &   5.30  &  0.11  & Sb    & SBN     &  $-$23.21 \\
1440$+$2511   &  0.0333  &  16.80$\pm$0.06  &  15.89$\pm$0.04  &  14.18$\pm$0.09  &  12.84$\pm$0.25   &    23$\pm$ 5  &  28.7   &   5.02  &  0.12  & Sb    & SBN     &  $-$23.00 \\
1440$+$2521S  &  0.0314  &  17.12$\pm$0.02  &  16.37$\pm$0.04  &  14.53$\pm$0.33  &  13.41$\pm$0.29   &    83$\pm$17  &  13.4   &   3.47  &  0.11  & Sb    & SBN     &  $-$22.52 \\
1442$+$2845   &  0.0110  &  15.53$\pm$0.02  &  14.85$\pm$0.03  &  12.97$\pm$0.10  &  11.90$\pm$0.09   &    81$\pm$16  &   8.2   &   4.82  &  0.07  & Sb    & SBN     &  $-$21.67 \\
1443$+$2714   &  0.0290  &  16.15$\pm$0.03  &  15.13$\pm$0.06  &  13.26$\pm$0.03  &  11.93$\pm$0.03   &   102$\pm$20  &  12.9   &   7.22  &  0.08  & Sa    & Sy2     &  $-$23.79 \\
1443$+$2844   &  0.0307  &  15.71$\pm$0.02  &  14.96$\pm$0.03  &  13.19$\pm$0.03  &  12.19$\pm$0.05   &    74$\pm$15  &  23.0   &   7.95  &  0.08  & SBc$+$  & SBN     &  $-$23.52 \\
1443$+$2548   &  0.0358  &  15.88$\pm$0.05  &  15.29$\pm$0.03  &  13.67$\pm$0.36  &  12.62$\pm$0.25   &    57$\pm$11  &  20.4   &   5.02  &  0.12  & Sc$+$   & SBN     &  $-$23.45 \\
1444$+$2923   &  0.0281  &  16.41$\pm$0.07  &  15.74$\pm$0.03  &  14.53$\pm$0.15  &  13.56$\pm$0.23   &    22$\pm$ 4  &  49.2   &   3.90  &  0.06  & S0    & DANS    &  $-$21.90 \\
1452$+$2754   &  0.0339  &  16.49$\pm$0.03  &  15.54$\pm$0.04  &  13.09$\pm$0.36  &  12.10$\pm$0.25   &    77$\pm$15  &  18.0   &   3.80  &  0.10  & Sb    & SBN     &  $-$23.90 \\
1506$+$1922   &  0.0205  &  16.07$\pm$0.04  &  15.01$\pm$0.04  &  12.90$\pm$0.37  &  11.97$\pm$0.26   &    78$\pm$16  &  19.5   &   3.91  &  0.14  & Sb    & HIIH    &  $-$23.00 \\
1513$+$2012   &  0.0369  &  16.27$\pm$0.03  &  15.30$\pm$0.03  &  13.56$\pm$0.03  &  12.33$\pm$0.06   &   109$\pm$22  &  14.5   &   4.56  &  0.12  & Sa    & SBN     &  $-$24.05 \\
1537$+$2506N  &  0.0229  &  15.21$\pm$0.02  &  14.30$\pm$0.03  &  12.24$\pm$0.07  &  11.27$\pm$0.07   &   113$\pm$22  &  27.2   &   3.90  &  0.15  & SBb   & HIIH    &  $-$23.75 \\
1537$+$2506S  &  0.0229  &  16.41$\pm$0.02  &  15.66$\pm$0.03  &  13.82$\pm$0.06  &  12.80$\pm$0.06   &   151$\pm$30  &   9.5   &   3.46  &  0.15  & SBa   & HIIH    &  $-$22.29 \\
\hline
\hline
\end{tabular}\\
\end{table*}\tabcolsep=0.1cm
\setcounter{table}{1}\begin{table*}
\caption{continued}
\begin{tabular}{lcrrrrrrrrllr} 
\hline 
UCM name    &    z     &     \multicolumn{1}{c}{$m_B$}      &      \multicolumn{1}{c}{$m_r$}     &     \multicolumn{1}{c}{$m_J$}     &      \multicolumn{1}{c}{$m_K$}     &$EW(\mathrm H\alpha)$&$3d_L$ (kpc)&  $\frac{F_{\mathrm H\alpha}}{F_{\mathrm H\beta}}$ & $A_V^{\mathrm Gal}$ &  MphT  &  SpT   &  $M_K$\\ 
 \multicolumn{1}{c}{(1)} & \multicolumn{1}{c}{(2)} & \multicolumn{1}{c}{(3)} & \multicolumn{1}{c}{(4)} & \multicolumn{1}{c}{(5)} & \multicolumn{1}{c}{(6)} & \multicolumn{1}{c}{(7)} &\multicolumn{1}{c}{(8)} & (9) & (10) & (11) & (12) & (13) \\
\hline 
\hline
1557$+$1423   &  0.0375  &  16.89$\pm$0.03  &  15.91$\pm$0.03  &  14.05$\pm$0.08  &  12.98$\pm$0.06   &    40$\pm$ 8  &  16.3   &   3.58  &  0.17  & Sb    & SBN     &  $-$23.20 \\
1612$+$1308   &  0.0114  &  18.66$\pm$0.02  &  17.75$\pm$0.03  &  16.88$\pm$0.06  &  15.97$\pm$0.18   &   510$\pm$102  &   2.5   &   2.89  &  0.16  & BCD   & BCD     &  $-$17.64 \\
1646$+$2725   &  0.0339  &  18.42$\pm$0.03  &  17.90$\pm$0.07  &  16.32$\pm$0.09  &  15.36$\pm$0.12   &   214$\pm$43  &  11.9   &   3.70  &  0.29  & Sc$+$   & DHIIH   &  $-$20.81 \\
1647$+$2950   &  0.0290  &  15.59$\pm$0.03  &  14.88$\pm$0.03  &  12.97$\pm$0.32  &  12.11$\pm$0.29   &    75$\pm$15  &  19.4   &   5.25  &  0.16  & Sc$+$   & SBN     &  $-$23.47 \\
1647$+$2729   &  0.0366  &  16.07$\pm$0.11  &  15.37$\pm$0.03  &  13.45$\pm$0.08  &  12.35$\pm$0.05   &    45$\pm$ 9  &  20.7   &   5.45  &  0.26  & Sb    & SBN     &  $-$23.76 \\
1647$+$2727   &  0.0369  &  16.10$\pm$0.05  &  16.57$\pm$0.03  &  14.91$\pm$0.04  &  13.95$\pm$0.06   &    56$\pm$11  &   7.2   &   4.76  &  0.28  & Sb    & SBN     &  $-$22.33 \\
1648$+$2855   &  0.0308  &  15.69$\pm$0.03  &  15.17$\pm$0.03  &  13.95$\pm$0.04  &  12.78$\pm$0.08   &   203$\pm$41  &  12.8   &   3.38  &  0.17  & Sa    & HIIH    &  $-$23.06 \\
1653$+$2644   &  0.0346  &  14.88$\pm$0.03  &   \multicolumn{1}{c}{$-$}  &  11.91$\pm$0.04  &  10.93$\pm$0.06   &     6$\pm$ 1  &  14.2   &  10.17  &  0.24  & INTER & SBN     &  $-$25.03 \\
1654$+$2812   &  0.0348  &  18.25$\pm$0.12  &  17.43$\pm$0.04  &  15.91$\pm$0.11  &  15.07$\pm$0.15   &    61$\pm$12  &  16.8   &   3.53  &  0.20  & Sc$+$   & DHIIH   &  $-$20.98 \\
1655$+$2755   &  0.0349  &  15.72$\pm$0.03  &  14.35$\pm$0.04  &  12.22$\pm$0.05  &  11.32$\pm$0.06   &    46$\pm$ 9  &  51.5   &   4.55  &  0.21  & Sc$+$   & Sy2     &  $-$24.63 \\
1656$+$2744   &  0.0330  &  17.73$\pm$0.02  &  16.45$\pm$0.20  &  14.50$\pm$0.11  &  13.25$\pm$0.08   &    69$\pm$14  &  12.1   &   4.51  &  0.33  & Sa    & SBN     &  $-$22.71 \\
1657$+$2901   &  0.0317  &  17.32$\pm$0.02  &  16.62$\pm$0.03  &  15.00$\pm$0.06  &  13.68$\pm$0.06   &    59$\pm$12  &   8.7   &   4.29  &  0.14  & Sb    & DANS    &  $-$22.31 \\
1659$+$2928   &  0.0369  &  15.78$\pm$0.05  &  14.78$\pm$0.04  &  12.80$\pm$0.07  &  11.73$\pm$0.08   &   154$\pm$31  &  71.2   &   4.23  &  0.16  & SB0   & Sy1     &  $-$24.36 \\
1701$+$3131   &  0.0345  &  15.33$\pm$0.02  &  13.70$\pm$0.03  &  12.46$\pm$0.06  &  11.48$\pm$0.07   &    45$\pm$ 9  &  43.7   &   9.89  &  0.10  & S0    & Sy1     &  $-$24.46 \\
2238$+$2308   &  0.0236  &  14.86$\pm$0.05  &  13.98$\pm$0.03  &  12.10$\pm$0.07  &  11.05$\pm$0.06   &    50$\pm$10  &  28.7   &   6.42  &  0.20  & Sa(r) & SBN     &  $-$24.05 \\
2239$+$1959   &  0.0237  &  15.05$\pm$0.01  &  14.26$\pm$0.03  &  12.57$\pm$0.07  &  11.48$\pm$0.04   &   118$\pm$24  &  17.8   &   4.65  &  0.16  & S0    & HIIH    &  $-$23.66 \\
2249$+$2149   &  0.0462  &  16.03$\pm$0.02  &  14.81$\pm$0.03  &  12.53$\pm$0.04  &  11.71$\pm$0.05   &     6$\pm$ 1  &  45.2   &   8.96  &  0.28  & Sb    & SBN     &  $-$24.88 \\
2250$+$2427   &  0.0421  &  15.40$\pm$0.02  &  14.82$\pm$0.03  &  12.95$\pm$0.07  &  11.67$\pm$0.04   &   138$\pm$28  &  39.5   &   5.19  &  0.49  & Sa    & SBN     &  $-$24.77 \\
2251$+$2352   &  0.0267  &  16.62$\pm$0.01  &  15.95$\pm$0.03  &  14.40$\pm$0.07  &  13.37$\pm$0.04   &    68$\pm$14  &   7.4   &   3.05  &  0.23  & Sc$+$   & DANS    &  $-$22.18 \\
2253$+$2219   &  0.0242  &  16.31$\pm$0.01  &  15.61$\pm$0.03  &  13.59$\pm$0.07  &  12.42$\pm$0.04   &    63$\pm$13  &   9.4   &   4.25  &  0.18  & Sa    & SBN     &  $-$22.82 \\
2255$+$1930S  &  0.0192  &  16.20$\pm$0.01  &  15.66$\pm$0.03  &  13.80$\pm$0.07  &  12.75$\pm$0.04   &    47$\pm$ 9  &   7.4   &   3.93  &  0.19  & Sb    & SBN     &  $-$21.97 \\
2255$+$1930N  &  0.0189  &  15.92$\pm$0.01  &  14.83$\pm$0.03  &  12.84$\pm$0.07  &  11.68$\pm$0.04   &    68$\pm$14  &  13.6   &   5.30  &  0.19  & Sb    & SBN     &  $-$22.99 \\
2255$+$1926   &  0.0193  &  17.03$\pm$0.02  &  16.33$\pm$0.05  &  14.82$\pm$0.09  &  13.91$\pm$0.08   &    34$\pm$ 7  &  13.8   &   3.13  &  0.18  & Sb    & HIIH    &  $-$21.03 \\
2255$+$1654   &  0.0388  &  16.72$\pm$0.03  &  15.32$\pm$0.09  &  13.01$\pm$0.08  &  11.53$\pm$0.05   &    27$\pm$ 5  &  37.7   &   4.05  &  0.19  & Sc$+$   & SBN     &  $-$24.70 \\
2256$+$2001   &  0.0193  &  15.69$\pm$0.04  &  14.64$\pm$0.04  &  12.86$\pm$0.05  &  12.05$\pm$0.09   &    14$\pm$ 3  &  29.6   &   9.60  &  0.14  & Sc$+$   & DANS    &  $-$22.58 \\
2257$+$2438   &  0.0345  &  15.57$\pm$0.05  &  15.82$\pm$0.08  &  13.51$\pm$0.05  &  12.08$\pm$0.05   &   347$\pm$69  &  22.5   &   5.21  &  0.51  & S0    & Sy1     &  $-$23.89 \\
2257$+$1606   &  0.0339  &  16.49$\pm$0.13  &   \multicolumn{1}{c}{$-$}  &  13.52$\pm$0.04  &  12.43$\pm$0.05   &    21$\pm$ 4  &   5.7   &   4.05  &  0.22  & S0    & SBN     &  $-$23.52 \\
2258$+$1920   &  0.0220  &  15.79$\pm$0.03  &  15.57$\pm$0.03  &  13.51$\pm$0.08  &  12.51$\pm$0.05   &   144$\pm$29  &  12.1   &   3.42  &  0.21  & Sc$+$   & DANS    &  $-$22.64 \\
2300$+$2015   &  0.0346  &  16.83$\pm$0.03  &  15.93$\pm$0.03  &  13.87$\pm$0.08  &  12.75$\pm$0.05   &    63$\pm$13  &  15.8   &   5.29  &  0.56  & Sb    & SBN     &  $-$23.33 \\
2302$+$2053W  &  0.0328  &  18.04$\pm$0.06  &  17.12$\pm$0.05  &  15.37$\pm$0.08  &  14.34$\pm$0.06   &   206$\pm$41  &  13.1   &   4.47  &  1.15  & Sb    & HIIH    &  $-$21.67 \\
2302$+$2053E  &  0.0328  &  15.85$\pm$0.05  &  14.58$\pm$0.03  &  12.81$\pm$0.08  &  11.64$\pm$0.05   &    26$\pm$ 5  &  20.2   &   6.73  &  1.14  & Sb    & SBN     &  $-$24.39 \\
2303$+$1856   &  0.0276  &  16.12$\pm$0.03  &  15.06$\pm$0.04  &  12.58$\pm$0.11  &  11.40$\pm$0.08   &    47$\pm$ 9  &  15.3   &   7.95  &  0.42  & Sa    & SBN     &  $-$24.17 \\
2303$+$1702   &  0.0428  &  17.35$\pm$0.05  &  16.29$\pm$0.03  &  14.39$\pm$0.27  &  13.35$\pm$0.04   &    44$\pm$ 9  &  20.1   &   3.88  &  0.32  & Sc$+$   & Sy2     &  $-$23.12 \\
2304$+$1640   &  0.0179  &  17.89$\pm$0.03  &  17.31$\pm$0.04  &  16.08$\pm$0.11  &  15.09$\pm$0.10   &   151$\pm$30  &   6.5   &   3.78  &  0.36  & BCD   & BCD     &  $-$19.57 \\
2304$+$1621   &  0.0384  &  17.14$\pm$0.03  &  15.42$\pm$0.04  &  14.04$\pm$0.26  &  13.04$\pm$0.04   &    48$\pm$10  &   7.7   &   3.77  &  0.42  & Sa    & DANS    &  $-$23.15 \\
2307$+$1947   &  0.0271  &  16.94$\pm$0.03  &  15.94$\pm$0.08  &  13.77$\pm$0.11  &  12.57$\pm$0.08   &    30$\pm$ 6  &  10.6   &   3.49  &  0.71  & Sb    & DANS    &  $-$23.08 \\
2310$+$1800   &  0.0363  &  16.89$\pm$0.03  &  15.83$\pm$0.03  &  13.55$\pm$0.11  &  12.32$\pm$0.08   &    41$\pm$ 8  &  18.6   &   5.81  &  0.56  & Sb    & SBN     &  $-$23.93 \\
2312$+$2204   &  0.0327  &  17.14$\pm$0.04  &   \multicolumn{1}{c}{$-$}  &   \multicolumn{1}{c}{$-$}  &  13.10$\pm$0.03   &    47$\pm$ 9  &   5.4   &   5.51  &  0.67  & Sa    & SBN     &  $-$22.83 \\
2313$+$1841   &  0.0300  &  17.19$\pm$0.09  &  16.25$\pm$0.03  &  14.28$\pm$0.11  &  13.09$\pm$0.10   &    60$\pm$12  &  15.8   &   6.15  &  0.42  & Sb    & SBN     &  $-$22.59 \\
2313$+$2517   &  0.0273  &  15.00$\pm$0.03  &   \multicolumn{1}{c}{$-$}  &  11.78$\pm$0.04  &  10.51$\pm$0.04   &    28$\pm$ 6  &  12.9   &   6.21  &  0.28  & Sa    & SBN     &  $-$24.96 \\
2315$+$1923   &  0.0385  &  17.55$\pm$0.03  &  16.98$\pm$0.03  &  15.50$\pm$0.06  &  14.65$\pm$0.07   &   164$\pm$33  &  14.9   &   4.62  &  0.23  & Sb    & HIIH    &  $-$21.54 \\
2316$+$2457   &  0.0277  &  14.62$\pm$0.03  &  13.63$\pm$0.06  &  11.72$\pm$0.11  &  10.49$\pm$0.08   &    35$\pm$ 7  &  24.6   &   4.85  &  0.34  & SBa   & SBN     &  $-$25.05 \\
2316$+$2459   &  0.0274  &  16.13$\pm$0.04  &  15.13$\pm$0.04  &  12.91$\pm$0.11  &  11.91$\pm$0.09   &    33$\pm$ 7  &  26.6   &   7.72  &  0.34  & Sc$+$   & SBN     &  $-$23.58 \\
2316$+$2028   &  0.0263  &  17.11$\pm$0.03  &  16.85$\pm$0.03  &  14.08$\pm$0.11  &  12.94$\pm$0.09   &    82$\pm$16  &   9.2   &   5.59  &  0.49  & Sa    & DANS    &  $-$22.61 \\
2317$+$2356   &  0.0334  &  14.16$\pm$0.10  &  13.35$\pm$0.03  &  11.43$\pm$0.04  &  10.55$\pm$0.05   &    28$\pm$ 6  &  36.2   &   8.54  &  0.25  & Sa    & SBN     &  $-$25.35 \\
2319$+$2234   &  0.0364  &  16.80$\pm$0.05  &  16.55$\pm$0.03  &  13.98$\pm$0.11  &  12.85$\pm$0.08   &    81$\pm$16  &  17.6   &   4.85  &  0.20  & Sb    & SBN     &  $-$23.25 \\
2319$+$2243   &  0.0313  &  15.82$\pm$0.10  &  14.76$\pm$0.03  &  12.78$\pm$0.05  &  11.77$\pm$0.04   &    34$\pm$ 7  &  26.3   &   8.37  &  0.23  & S0    & SBN     &  $-$23.94 \\
2320$+$2428   &  0.0328  &  15.89$\pm$0.05  &  14.60$\pm$0.03  &  12.33$\pm$0.04  &  11.08$\pm$0.02   &     9$\pm$ 2  &  28.9   &   9.27  &  0.21  & Sa    & DANS    &  $-$24.79 \\
2321$+$2149   &  0.0374  &  16.66$\pm$0.04  &  16.02$\pm$0.03  &  14.28$\pm$0.11  &  13.30$\pm$0.08   &    53$\pm$11  &  17.9   &   4.20  &  0.22  & Sc$+$   & SBN     &  $-$22.91 \\
2321$+$2506   &  0.0331  &  15.79$\pm$0.04  &  15.33$\pm$0.04  &  13.70$\pm$0.05  &  12.73$\pm$0.06   &    43$\pm$ 9  &  25.2   &  10.32  &  0.17  & Sc$+$   & SBN     &  $-$23.10 \\
2322$+$2218   &  0.0249  &  17.77$\pm$0.02  &  16.59$\pm$0.08  &  14.39$\pm$0.04  &  13.25$\pm$0.02   &    41$\pm$ 8  &  10.0   &   5.70  &  0.15  & Sc$+$   & SBN     &  $-$22.02 \\
2324$+$2448   &  0.0123  &  13.59$\pm$0.04  &  12.80$\pm$0.03  &  10.52$\pm$0.11  &   9.54$\pm$0.08   &     9$\pm$ 2  &  20.3   &   4.57  &  0.23  & Sb    & SBN     &  $-$24.16 \\
2325$+$2318   &  0.0114  &  13.28$\pm$0.04  &   \multicolumn{1}{c}{$-$}  &   \multicolumn{1}{c}{$-$}  &  10.55$\pm$0.04   &    87$\pm$17  &   8.7   &   4.21  &  0.14  & INTER & HIIH    &  $-$22.93 \\
2325$+$2208   &  0.0116  &  12.59$\pm$0.05  &  11.81$\pm$0.04  &  10.16$\pm$0.08  &   9.06$\pm$0.07   &    36$\pm$ 7  &  47.4   &   9.43  &  0.16  & SBc$+$  & SBN     &  $-$24.45 \\
2326$+$2435   &  0.0174  &  16.61$\pm$0.02  &  16.03$\pm$0.03  &  14.61$\pm$0.06  &  13.77$\pm$0.09   &   211$\pm$42  &  12.5   &   3.66  &  0.33  & Sb    & DHIIH   &  $-$20.70 \\
2327$+$2515N  &  0.0206  &  15.79$\pm$0.03  &  15.45$\pm$0.03  &  14.14$\pm$0.10  &  13.24$\pm$0.12   &    94$\pm$19  &   9.1   &   3.71  &  0.20  & Sb    & HIIH    &  $-$21.65 \\
2327$+$2515S  &  0.0206  &  15.80$\pm$0.03  &  15.23$\pm$0.03  &  13.95$\pm$0.10  &  13.06$\pm$0.13   &   257$\pm$51  &  11.7   &   4.56  &  0.20  & S0    & HIIH    &  $-$21.88 \\
2329$+$2427   &  0.0200  &  15.92$\pm$0.05  &  14.68$\pm$0.03  &  12.62$\pm$0.05  &  11.51$\pm$0.03   &    13$\pm$ 3  &  23.4   &   9.87  &  0.30  & Sb    & DANS    &  $-$23.23 \\
2329$+$2500   &  0.0305  &  16.11$\pm$0.04  &  15.28$\pm$0.04  &  13.24$\pm$0.18  &  12.20$\pm$0.04   &   180$\pm$36  &  26.5   &   4.54  &  0.22  & S0(r) & Sy1     &  $-$23.49 \\
2329$+$2512   &  0.0133  &  16.88$\pm$0.02  &  16.28$\pm$0.03  &  14.78$\pm$0.04  &  14.08$\pm$0.05   &    58$\pm$12  &   4.9   &   3.81  &  0.15  & Sa    & DHIIH   &  $-$19.78 \\
2331$+$2214   &  0.0352  &  17.75$\pm$0.04  &  16.57$\pm$0.03  &  14.67$\pm$0.04  &  13.59$\pm$0.04   &    60$\pm$12  &  12.8   &   5.82  &  0.20  & Sb    & SBN     &  $-$22.38 \\
2333$+$2248   &  0.0399  &  16.97$\pm$0.03  &  16.31$\pm$0.08  &  14.70$\pm$0.06  &  13.74$\pm$1.23   &   177$\pm$36  &  56.6   &   4.08  &  0.22  & Sc$+$   & HIIH    &  $-$22.51 \\
2333$+$2359   &  0.0395  &  17.20$\pm$0.04  &  16.02$\pm$0.03  &  14.03$\pm$0.14  &  12.79$\pm$0.03   &    51$\pm$10  &  13.3   &   3.45  &  0.26  & S0a   & Sy1     &  $-$23.59 \\
\hline
\hline
\end{tabular}\\
\end{table*}\tabcolsep=0.1cm
\setcounter{table}{1}\begin{table*}
\caption{continued}
\begin{tabular}{lcrrrrrrrrllr} 
\hline 
UCM name    &    z     &     \multicolumn{1}{c}{$m_B$}      &      \multicolumn{1}{c}{$m_r$}     &     \multicolumn{1}{c}{$m_J$}     &      \multicolumn{1}{c}{$m_K$}     &$EW(\mathrm H\alpha)$&$3d_L$ (kpc)&  $\frac{F_{\mathrm H\alpha}}{F_{\mathrm H\beta}}$ & $A_V^{\mathrm Gal}$ &  MphT  &  SpT   &  $M_K$\\ 
 \multicolumn{1}{c}{(1)} & \multicolumn{1}{c}{(2)} & \multicolumn{1}{c}{(3)} & \multicolumn{1}{c}{(4)} & \multicolumn{1}{c}{(5)} & \multicolumn{1}{c}{(6)} & \multicolumn{1}{c}{(7)} &\multicolumn{1}{c}{(8)} & (9) & (10) & (11) & (12) & (13) \\
\hline 
\hline
2348$+$2407   &  0.0359  &  17.09$\pm$0.04  &  16.43$\pm$0.03  &  14.61$\pm$0.05  &  13.60$\pm$0.05   &    56$\pm$11  &  21.5   &   4.10  &  0.22  & Sa    & SBN     &  $-$22.46 \\
2351$+$2321   &  0.0273  &  17.77$\pm$0.02  &  16.44$\pm$0.05  &  14.94$\pm$0.07  &  13.94$\pm$0.06   &    92$\pm$18  &  16.6   &   2.86  &  0.31  & Sb    & HIIH    &  $-$21.51 \\
\hline
\hline
\label{input}
\end{tabular}
\setcounter{table}{1}
\caption{Photometric and spectroscopic data for the 191 UCM Survey galaxies. Columns stand for: (1) UCM name established in \citet{1994ApJS...95..387Z,1996ApJS..105..343Z}. (2) Redshift \citep{1996A&AS..120..323G}. (3)$-$(6) Johnson $B$, Gunn $r$, $J$ and $K$ magnitudes and errors at three disk$-$scales measured in $r$. (7) $\mathrm H\alpha$ equivalent width \citep{1996A&AS..120..323G}. (8) Disk scale (as explained in the main text) in kpc. (9) Intensity ratio between the $\mathrm H\alpha$ and $\mathrm H\beta$ lines corrected for stellar absorption (see text). (10) Galactic $V$$-$band extinction \citep{1998ApJ...500..525S}. (11) Morphological type \citep{2001A&A...365..370P}. (12) Spectroscopic type \citep{1996A&AS..120..323G}. (13) Absolute $K$$-$band magnitude corrected for Galactic extinction.}
\end{table*}

\subsection{Archival data}

\begin{figure}
\center{\psfig{file=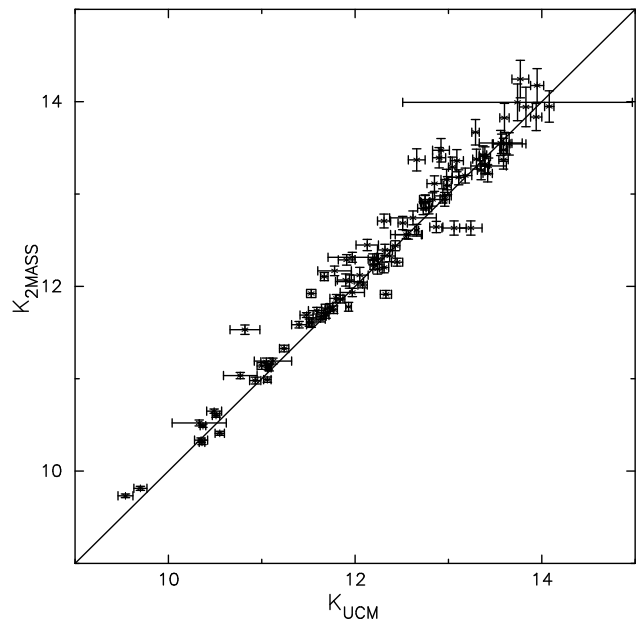}}
\caption{Photometry comparison of the $K$ band total magnitudes for the
UCM Survey galaxies included in the 2 Micron All Sky Survey, Second
Incremental Release.}
\label{2MASSvsUCM}
\end{figure}

At the time of writing, a total of 97 galaxies in our sample have been
observed in $J$ and $K$ as part of the Two Micron All Sky Survey
\citep[2MASS; for details on the source identification and photometry procedures see][]
{2000AJ....119.2498J}.  When we compare our total magnitudes with the
total magnitudes derived by the 2MASS team, we find that the 2MASS
total magnitudes are, on average, 0.07 mag fainter than ours both in
$J$ and $K$. The largest differences are mostly found in objects
showing companions or field stars. This offset is probably due to
differences in the determination of the total magnitudes. Indeed, when
we compare the magnitudes inside the same aperture in both the 2MASS
images and in ours, we find that the average differences (weighted
with the photometric errors) are $0.001\pm0.052~{\mathrm mags}$ in $K$ and
$0.003\pm0.038~{\mathrm mags}$ in $J$. Fig.~\ref{2MASSvsUCM} shows the
comparison in the $K$ band.

Among the 97 galaxies common to both samples, a total of 20 objects in
$J$ and 5 in $K$ have not been imaged by us or our data are of poor
quality.  For these galaxies we have used the 2MASS images and
determined aperture and total magnitudes following the procedures
described in Section~\ref{photo}.  These magnitudes will be used in
our analysis.

\subsection{Summary of available data}

Table~\ref{input} contains all the data described in this section. It
includes object names, redshifts, magnitudes, and errors in the four
photometric bands, together with $\mathrm H\alpha$ equivalent widths and
uncertainties, radii of the apertures used in the photometric
measurements, $\mathrm H\alpha/\mathrm H\beta$ intensity ratios, Galactic
extinction values in the $V$ band, morphological and spectroscopic
types and total absolute $K$ magnitudes.

Before attempting the comparison with the models, the $BrJK$ magnitudes listed
in Table~\ref{input} were corrected for Galactic extinction using the maps of
\citet{1998ApJ...500..525S} and the extinction curve of
\citet{1989ApJ...345..245C}. We also applied k-corrections given by
\citet{1999A&A...351..869F} for $BJK$ and \citet{1995PASP..107..945F} for
Gunn-$r$, taking into account the morphological types. The k-corrections
applied are, in any case, small because of the low redshifts of the galaxies in
the sample ($z<0.045$). The k-corrections are 
(in absolute value) smaller than  0.22 in $B$, 0.04
in $r$, 0.03 in $J$ and 0.13 in $K$. Note that the nIR k-corrections are
negative.

\section{Models}
\label{method}

\subsection{Underlying stellar population}

In our models, we have assumed that our observational data ($B-r$,
$r-J$, and $J-K$ colours, and $EW(\mathrm H\alpha)$) can be reproduced by an
underlying stellar population with colours and mass-to-light ratios in
the $K$-band ($M/L_K$ hereafter) similar to those of typical spiral
and lenticular galaxies of the same morphological type on top of which
a recent burst of star formation is superimposed. This assumption
represents a significant improvement with respect to \gil\, where the
same underlying population colours and $M/L_K$ were assumed for the
entire sample. We have also considered typical values for the EW of
the $\mathrm H\alpha$ emission-line in `normal' spirals
\citep{1992AJ....103.1512D,1983ApJ...272...54K}. This fact means that
our modelling will refer to the properties of a recent star formation
event which takes place {\it in excess of what is typical in a normal
spiral or lenticular galaxy}. In Table~\ref{oldpop} we give the
typical $B-r$ \citep{1995PASP..107..945F}, $r-J$, and $J-K$ colours
\citep{1999A&A...351..869F}, $EW(\mathrm H\alpha)$,  and $M/L_K$ for each 
morphological type. The $M/L_K$ values have been derived separately
for each galaxy type and IMF using a relation between the $B-r$ color
and the $M/L_K$ \citep[see][]{2000MNRAS.312..497B,2001ApJ...550..212B}
for the Bruzual \& Charlot (private communication; BC99 hereafter)
exponential star formation models with different $\tau$ parameters, a
formation age of 12\,Gyr, and a mean attenuation in the $V$-band of
$\tau_{V,\mathrm{ISM}}=0.5\,\mathrm mag$.

With regard to the Blue Compact Dwarf galaxies there is a significant
lack of studies providing information about the optical and nIR
properties of their underlying stellar population. Despite of the
recent efforts, both at optical \citep{2001ApJS..133..321C} and nIR
wavelengths \citep{2001A&A...367...33D}, very few objects have been
studied simultaneously within the wavelength range defined by the $B$
and $K$ bands. A noteworthy exception is the work of
\citet{2000A&A...361..465G,2000A&AS..145..377G} on
the BCD galaxy Mrk~86 where deep surface photometry was obtained in
all $BVRJHK$ bands. It is important to note that this galaxy is a
prototype of the iE BCDs \citep{1986sfdg.conf...73L}, the most
numerous BCD subclass
\citep{1996A&AS..120..207P,2001ApJS..133..321C}. Moreover, the $B-R$
and $J-K$ colours of the underlying stellar population in Mrk~86
($B-R$=1.2; $J-K$=1.1; see Table~\ref{oldpop}) are very similar to the
average values derived by \citet{2001ApJS..133..321C}, $B-R$=1.1, and
\citet{2001A&A...367...33D}, $J-K$=1.0. The standard deviations of 
these mean values are 0.2\,mag in both cases.

Although there are no galaxies in our sample morphologically
classified as ellipticals, we also give the typical colours of this
type for the sake of completeness. These underlying population colours
are quite similar to our measurements in the outer parts of some
randomly selected test galaxies \citep{2002ApJnotyetii}.

\begin{table}
\caption{Assumed properties of the underlying stellar populations.}
\label{oldpop}
\tabcolsep=0.04cm
\begin{tabular}[c]{lccccccc}
\hline
Type$^1$&$(B-r)^2$&$(r-J)^3$&$(J-K)^4$ &$EW^5$&      &$M/L_K$&  \\
        &         &         &          &    (\AA)       &SALP$^6$&SCA$^7$&MSCA$^8$ \\
\hline				                       			     
\hline				                       			     
   E    &   1.15  &   1.90  &  0.91    &      0         & 1.24 &  0.65 & 0.55 \\
   S0   &   0.98  &   2.03  &  0.94    &     -2         & 1.01 &  0.57 & 0.43 \\
   Sa   &   0.92  &   1.92  &  1.01    &      0         & 0.95 &  0.54 & 0.40 \\
   Sb   &   0.69  &   2.07  &  1.01    &      8         & 0.73 &  0.45 & 0.30 \\
   Sc+  &   0.61  &   1.91  &  0.93    &     15         & 0.67 &  0.42 & 0.27 \\
   Irr  &   0.61  &   1.62  &  0.93    &     18         & 0.67 &  0.42 & 0.27 \\
   BCD  &   0.83  &   1.77  &  1.06    &     -2         & 0.86 &  0.51 & 0.36 \\
\hline
\hline
\end{tabular}
\setcounter{table}{2}
\caption{Main properties of the underlying population assumed in our 
models as a function of Hubble type (column 1). $B-r$, $r-J$, $J-K$
colours (columns 2,3 and 4), $\mathrm H\alpha$ equivalent width (column 5;
minus sign means absorption) and mass-to-light ratio in the $K$-band
for different IMFs (Salpeter, Scalo and Miller-Scalo in columns 6,7
and 8, respectively).}
\end{table}

Because the detection limit in $EW(\mathrm H\alpha)$ for the UCM Survey is
about 20\,\AA\ \citep{1995ApJ...455L...1G}, even late-type spirals
galaxies in the sample must have, or have recently had, enhanced star
formation compared to their `relaxed' counterparts in order to have
been detected in the UCM Survey photographic plates. The primary goal
of this paper will be the characterization of this star formation
activity.

\subsection{Recent star formation}
\label{mod}

In order to reproduce the observational properties of the sample we have
generated a complete set of models that assume a recent/ongoing  episode of
star formation that takes place in galaxies with the underlying stellar
population described above. For the stellar continuum of the newly-formed
stars, we use the predictions given by two different evolutionary synthesis
models developed by  BC99 and \citet[\SB\, hereafter]{1999ApJS..123....3L}.
Each of them allows to choose different star formation histories, IMFs and
metallicities.

From the number of Lyman photons predicted by these models, we have
computed the nebular continuum contribution using the emission and
recombination coefficients given by
\citet{1980PASP...92..596F} for $T_{\mathrm{e}}=10^{4}$\,K. For the
Balmer, Paschen, and Brackett hydrogen recombination-lines, 
luminosities (and the corresponding equivalent widths) have been
derived assuming the relation given by
\citet{1971MNRAS.153..471B} and the theoretical line-ratios expected for a 
low density gas ($n_{\mathrm{e}}=10^{2}\,\mathrm cm^{-3}$) with
$T_{\mathrm{e}}=10^{4}\,\mathrm K$ in Case B recombination
\citep{1989agna.book.....O}. Our values of 
the nebular continuum luminosity are systematically a $\sim15$\%
larger than the ones given by the \SB\, models, probably due to
differences in the assumed emission coefficients. The contribution of
the most intense forbidden emission-lines ($\mathrm{[O{
II}]\lambda\lambda3726,3729}$\,\AA, $\mathrm{[O{
III}]\lambda\lambda4959,5007}$\,\AA, $\mathrm{[N{
II}]\lambda\lambda6548,6583}$\,\AA, $\mathrm{[S{
II}]\lambda\lambda6717,6731}$\,\AA) to the bandpasses under study has
been also determined assuming the mean line ratios given by
\citet{1996A&AS..120..323G} for the sample. 
Following a complementary method, \citet{2001MNRAS.323..887C} 
have calculated all these line intensities using a photoionization code
in order to establish stronger constrains on the inferred star 
formation rates. We have decided not to follow their  approach 
since it would introduce more model-dependent parameters and
complicate the interpretation of the results. 

The predictions for the young and underlying stellar populations have
been combined using the ratio between the stellar mass of the young
stellar population over the total stellar mass of the galaxy (i.e.,
the burst strength, $b$) as a parameter.

\subsection{Recent star formation vs. old stellar population}

\begin{figure}
\center{\psfig{file=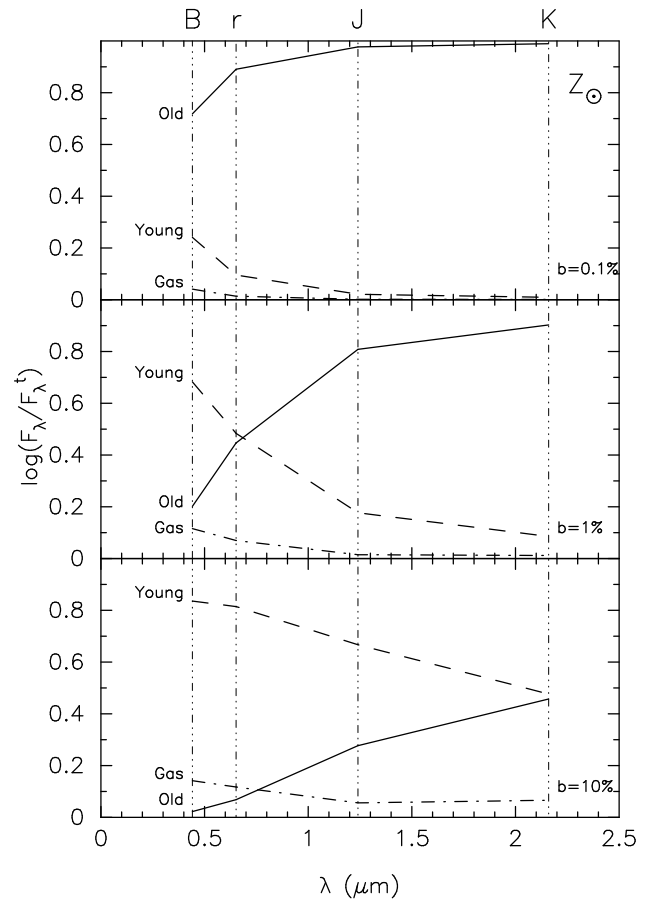}}
\caption{Comparison of the relative contribution of the older and younger 
populations and the gas to the total flux of our modelled galaxies as a
function of wavelength. The four photometric broad bands available for
the UCM sample are marked. Three cases are considered for an Sb galaxy
experiencing a recent ($5\,$ Myr) instantaneous burst with solar
metallicity and strengths 0.1\%, 1\% and 10\% of the total stellar
mass of the galaxy.}
\label{contrib}
\end{figure}

Fig.~\ref{contrib} depicts the relative importance of the three
sources of galaxy light considered in our models: young stars formed
in a recent burst, gas (continuum spectrum plus emission lines) and
the underlying evolved population. Each of the panels displays the
contribution of these sources to the total spectral energy
distribution of a typical Sb galaxy (whose colours are given in
Table~\ref{oldpop}), the most frequent Hubble type in the UCM
sample. This galaxy is experiencing a recent instantaneous burst with
a typical age of $\sim5\,$Myr (cf. \pii) and solar metallicity. Three
burst strengths have been considered: 0.1\%, 1\% and 10\% of the total
stellar mass. The four photometric bands available for our sample
($BrJK$) are marked.

This figure shows how important a recent burst of star formation can
be on the luminosity of a galaxy. A moderate burst of 1\% of the total
mass clearly dominates the blue optical spectrum. At longer
wavelengths, although the effect is reduced, the young stellar
population accounts for $\sim10$\% of the $K$-band luminosity. For a
stronger burst ($b=10$\%) the recent star formation contributes with
more than 80\% of the $B$-band light, and half of the total $K$-band
luminosity. This illustrates the need of a careful analysis of the
star formation history when determining stellar masses using optical
photometry, and, to a lesser extent, nIR data. We will come back to
this issue in \pii. We also remark the importance of the gaseous
contribution, mostly at optical wavelengths \citep[for a more detailed
discussion see][]{1995A&A...303...41K}.

\subsection{Dust attenuation}

\begin{figure*}
\center{\psfig{file=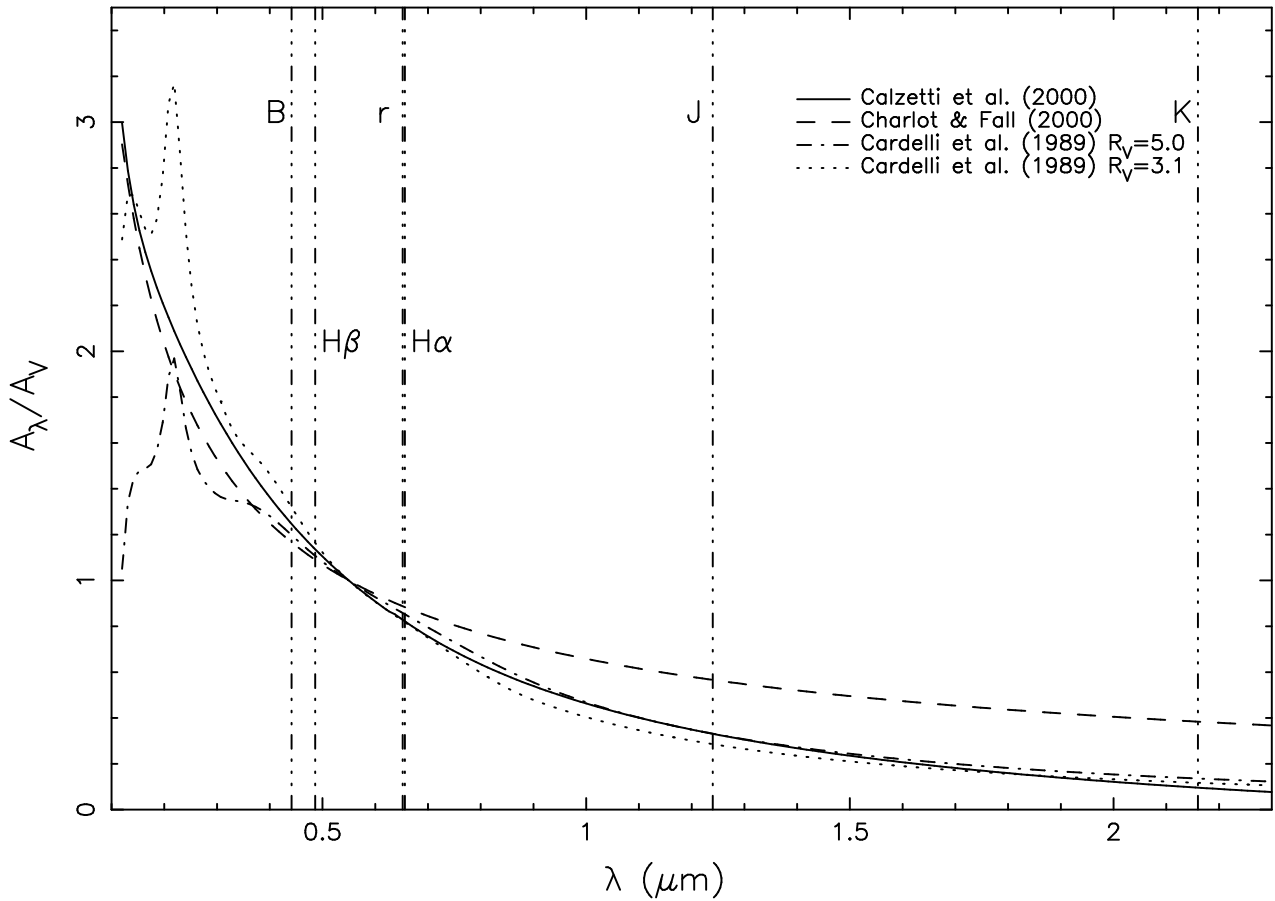}}
\caption{Wavelength dependence of 4 extinction laws: 
\citet{2000ApJ...533..682C}; \citet[][bursts ages younger than $10^7$ 
years have been assumed]{2000ApJ...539..718C} and
\citet{1989ApJ...345..245C} for R=3.1 and R=5.0. The effective
wavelengths of the bands considered in this work are shown. }
\label{atten}
\end{figure*}

Instead of correcting our observational data for internal extinction,
we decided to implement the reddening correction in our models when
predicting the optical-nIR colours and $EW(\mathrm H\alpha)$.  In
order to do so we have applied two alternative recipes, the one given
by \citet[\cf\, hereafter]{2000ApJ...539..718C}, and the one presented
by \citet[\calz\, from now on]{2000ApJ...533..682C}. These recipes
cope with three distinct problems: (1) the extinction law, i.e., the
wavelength dependence of the attenuation; (2) the differences between
the attenuation of the gas and the stellar emission; and (3) the
translation of these recipes into observables such as the colour
excess calculated with the Balmer decrement.

In the case of the \cf\, recipe we used the attenuation curve
parametrized by \calz\, instead of that given by these
authors. Although both attenuation curves are able to reproduce the
observational properties of starburst galaxies in the UV-optical
range, the one used in \cf\, leads to unrealistically low
optical-to-nIR colour excesses. In Fig.~\ref{atten} we show the
attenuation curves of \calz\,({\it solid line}),
\cf\, ({\it dashed-line}), and the Galactic extinction curve
\citep{1989ApJ...345..245C} for total-to-selective extinction ratios
($R_V$) of 3.1 ({\it dotted-line}) and 5.0 ({\it dash-dotted}). This
figure shows the attenuation law given by \cf\, for burst ages younger
than $10^7$ years, i.e., with power-law index $n=-0.7$. We have not
considered the effect of the finite lifetimes of the birth clouds
(explained in \cf) since the bursts in the UCM galaxies are rather
young (cf. \pii). \cf\,'s law is `too grey' at wavelengths longer than
the $r$-band. Therefore, we used the \calz\, attenuation curve also
for the \cf\, extinction recipe. This means that both recipes only
differ in how they relate the colour excess to the extinction of the
ionized gas, and this to the attenuation of the stellar
continuum. Each one of these issues are explained below.

The \cf\, recipe states that the stars in the burst are embedded in a
gaseous cloud with two layers, an internal HII region and a more
external HI envelope. This is immersed in the galaxy inter-stellar
medium. Given this scenario, \cf\, introduce a formulation for the
attenuation of the different components. Following their notation, the
attenuation of the ionized-gas emission can be written as
$(1-f)\times\tau_{\mathrm{BC}}+\tau_{\mathrm{ISM}}$, where
$\tau_{\mathrm{BC}}$ is the attenuation in the birth cloud associated
with the burst
($\tau_{\mathrm{BC}}=\tau_{\mathrm{HI}}+\tau_{\mathrm{HII}}$),
$\tau_{\mathrm{ISM}}$ is the attenuation due to the ISM, and $f$ is
the fraction of the attenuation in the birth cloud due to the HII
region (i.e. $f=\tau_{\mathrm{HII}}/\tau_{\mathrm{BC}}$).

Therefore, since the attenuation of the ionized-gas emission is known
from the $\mathrm H\alpha/\mathrm H\beta$ Balmer decrements given by
\citet{1996A&AS..120..323G} we can estimate the burst
($\tau_{\mathrm{BC}}+\tau_{\mathrm{ISM}}$) and underlying stellar
populations attenuations ($\tau_{\mathrm{ISM}}$) for a given $f$ and
$\tau_{V,\mathrm{ISM}}$. This method also deals with the extinction of
the emission-line flux. We have assumed $f=0.1$ and
$\tau_{V,\mathrm{ISM}}=0.5$, following \cf.  In the cases where the
calculated $\tau_{\mathrm{BC}}$ is incompatible with the measured
$E(B-V)_{\mathrm{gas}}$, the former was set to zero.

The extinction recipe given in \calz\, is empirical. It is based on
the comparison of fluxes in the UV and optical ranges for nearby
starburst galaxies. It considers that the stellar continuum flux is
affected by an effective extinction characterized by
$E(B-V)_{\mathrm{continuum}}$, which directly relates to the
measurable gas attenuation $E(B-V)_{\mathrm{gas}}$ via:
\begin{equation}
	E(B-V)_{\mathrm{continuum}}=0.44\cdot E(B-V)_{\mathrm{gas}}
\end{equation}

The recipe also includes the average attenuation law given in
Fig.~\ref{atten}.

\subsection{Fitting procedure}
\label{fitting}

In our analysis several `parameters' must be selected {\it a
priori}. These are:

	\begin{itemize}

		\item[--]The evolutionary synthesis model: BC99 or \SB.

		\item[--]The star-forming mode of the youngest stellar
		population: instantaneous or continuous star formation
		rate. These modes will be referred to as INST and
		CONS.

		\item[--]The IMF: \citet{1955ApJ...121..161S},
		\citet{1986FCPh...11....1S}, or
		\citet{1979ApJS...41..513M}. In all cases, we use
		$\mathcal{M}_{\mathrm{low}}=0.1\,\mathcal{M}_\odot$
		and $\mathcal{M}_{\mathrm{up}}=100\,\mathcal{M}_\odot$
		for the lower and upper mass limits of the IMF.

		\item[--]The extinction recipe: \cf\, or \calz.

	\end{itemize}

Once these have been fixed, the method leaves 3 free parameters
describing the newly-formed stars: (1) the age (from 0.89 to 100 Myr);
(2) metallicity of the burst ($1/5\,Z_{\odot}$, $2/5\,Z_{\odot}$,
$Z_{\odot}$, $2.5\,Z_{\odot}$, $5\,Z_{\odot}$), and (3) the burst
strength (from 0.01\% to 100\%).

The best-fitting model for each galaxy in the sample was derived using
the method described in \citet{2002AJ....123.1864G}. Briefly, this
procedure reproduces the Gaussian probability distributions associated
with the observational errors in $B-r$, $r-J$, $J-K$, and $2.5\cdot
\log[EW(\mathrm H\alpha)]$ using Monte Carlo simulations with a total of
1000 test `particles'. Comparing these particles with our models for
the range of parameters given above, we obtain a total of 1000
solutions. The comparison was carried out using a model grid
containing $\sim 2\cdot10^4$ points in the BC99 case and $\sim
2\cdot10^5$ for the \SB\, models. Both a reduced $\chi^{2}$ and a
Maximum Likelihood estimator were used to measure the goodness of the
fit. We included 2--3 colour terms and an $EW(\mathrm H\alpha)$
term. The observational uncertainties were taken into account. We used
the following formulae:
\begin{equation}
{\cal L}(t,b,Z) = \left(\prod_{n=1}^{3-4} {1\over \sqrt{2\pi}\Delta C_n}
 \exp\left( - {(c_n-C_n)^2\over 2 \Delta C_n\,^2}\right)\right)^{1/N}
\end{equation}
\begin{equation}
\chi^2=\frac{1}{N}\sum_{n=1}^{3-4} {\frac{(c_n-C_n)^2}{\Delta C_n^2}}
\end{equation}
where $C_n$ and $c_n$ are, respectively, the observed and modelled data
values (colours and $2.5\cdot \log\,EW(\mathrm H\alpha)$), $\Delta C_n$ are
their corresponding errors and N is the number of terms in the sum or
the product. $N=3$ ($N=4$) when we used two (three) colours plus
$EW(\mathrm H\alpha)$.

The distributions in the space of solutions were studied using
Principal Component Analysis. This fitting procedure gives the
best-fitting set of model parameters, the corresponding uncertainty
intervals, and the possible degeneracies between these parameters
within the uncertainty intervals. See \citet{2002AJ....123.1864G} for
details.

\section{Discussion}
\label{discussion}

\subsection{Goodness of the fit}

\begin{figure*}
\center{\psfig{file=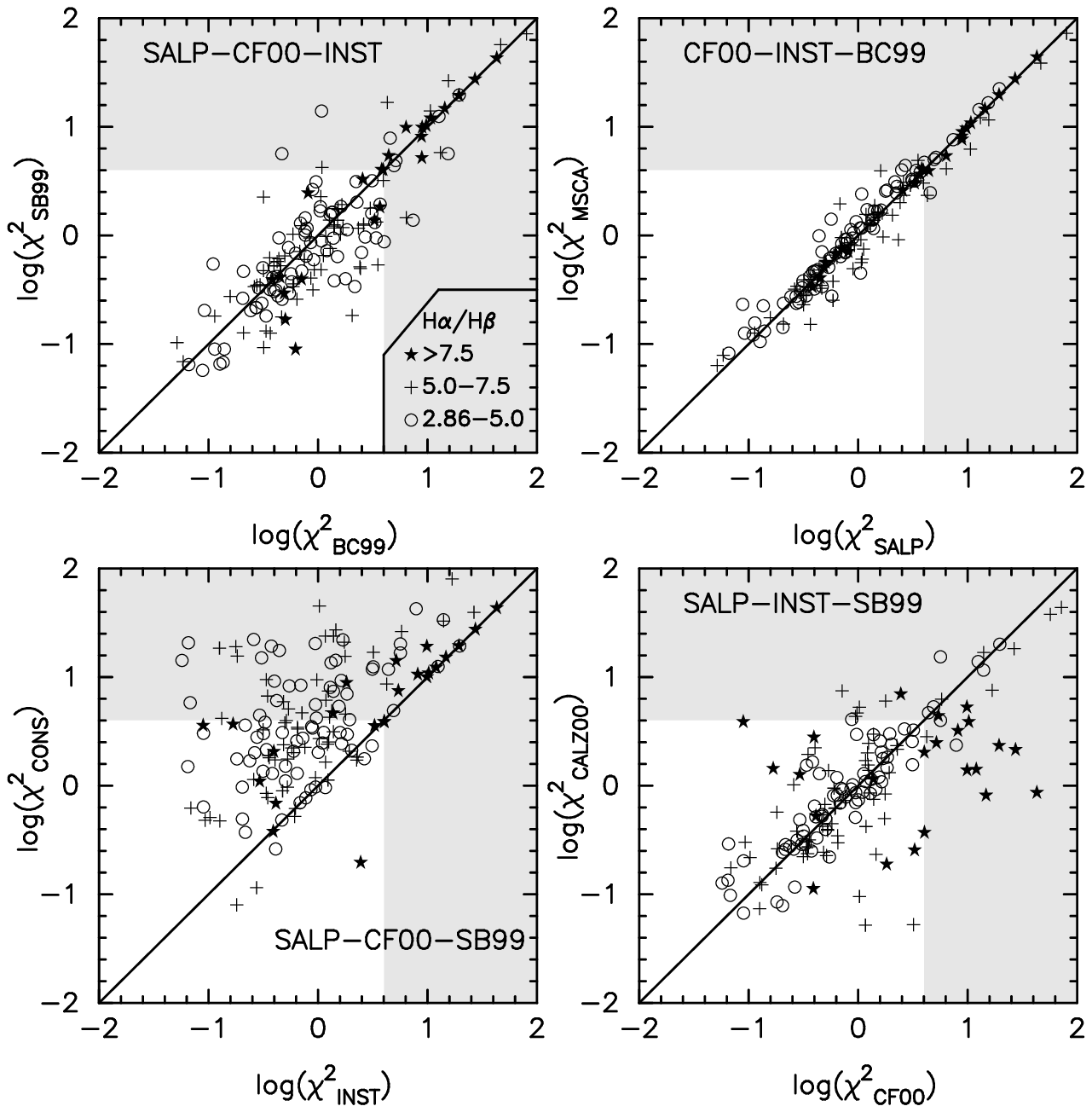,width=17cm}}
\caption{Plots of the $\chi^2$ obtained 
in the best-fittings comparing several {\it a priori} model
inputs. Different symbols represent different $\mathrm H\alpha/\mathrm H\beta$
line ratios (i.e., different extinctions). The top-left diagram
compares the two families of stellar synthesis models (BC99 and SB99)
for the same values of the other input parameters (i.e., Salpeter IMF,
\cf\, recipe and instantaneous SFR; see labels in the upper-left
corner). Different IMFs are compared in the upper-right diagram, star
formation scenarios in the bottom-left one and extinction recipes in
the bottom-right plot.}
\label{compchi}
\end{figure*}

\begin{figure*}
\center{\psfig{file=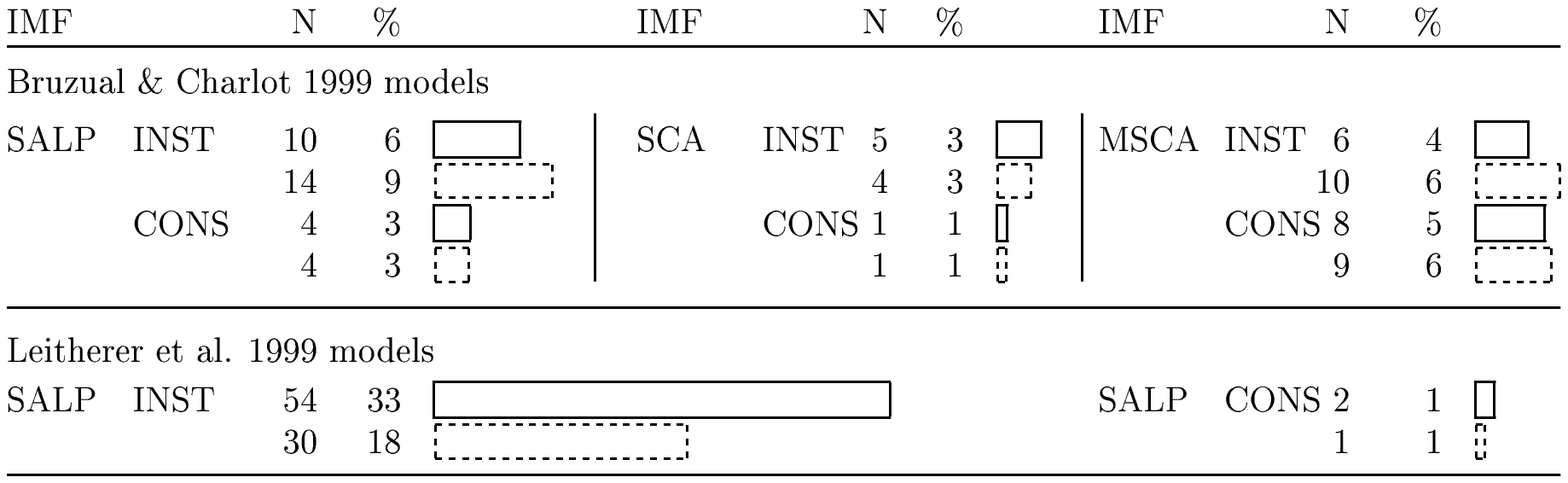,width=17cm}}
\caption{Distribution of the best-fittings (those with a lowest
value of $\chi^2$) for the UCM sample according to the input
parameters. Continuous-line rectangles stand for \cf\, extinction and
dashed ones for the \calz\, law (with sizes proportional to the total
number of objects).}
\label{bestone}
\end{figure*}

Somewhat surprisingly, we did not find significant differences in the results
obtained with the $\chi^2$ and the Maximum Likelihood estimators. Therefore,
all the following discussion (and the results for the fitted parameters given
in \pii) will refer to the modelling performed with the $\chi^2$
minimization. 

Out of the 163 UCM galaxies (excluding AGNs) with more than two
observed broad-bands, a total of 9 galaxies present $\chi^2$ values
greater than 4.0 {\it in all possible models considered}.  This
$\chi^2$ value corresponds to average differences between the observed
and modelled colours of $\sim0.3$\,mag ($\sim30$\% in flux) for typical
uncertainties of 0.15 mag in the colours and considering the EW term
as negligible. Two of these galaxies (UCM2304$+$1621 and
UCM2351$+$2321) present best-fittings which perfectly match the $B$,
$J$ and $K$ luminosities, but fail to reproduce the $r$-band
magnitudes by 0.3--0.5 mag, indicating that there may be a problem
with their $r$-band data.  Three of the remaining objects with high
$\chi^2$ values are face-on spirals with resolved structure
(UCM1304$+$2818, UCM2249$+$2149 and 2302$+$2053E), and another one
(UCM2255$+$1654) is an edge-on galaxy.  All of them exhibit strong dust
lanes, most visible in the $B$ band, that may indicate a complex
extinction behaviour (see discussion below). The remaining three
galaxies (UCM1647$+$2727, UCM1657$+$2901 and UCM2316$+$2028) are
compact objects that seem to have a burst affecting the whole galaxy
\citep[revealed by our $\mathrm H\alpha$ images,][]{2002ApJnotyet}.

The minimum number of rejected fits\footnote{Fits are rejected if
$\chi^2>4$} (19 galaxies) is achieved for \SB\, models with an
instantaneous burst, Salpeter IMF and \calz\, extinction.  Using the
same parameters, 20 rejected fits were found for BC99 models. In other
model/parameter combinations, the number of rejected fits increases.
For example, 26 fits are rejected with \SB, instantaneous burst,
Salpeter IMF and the \cf\, recipe.  Up to 74 are rejected for
continuous SFR models.  All the objects without valid fits will not be
used in the following discussion. We have kept the two galaxies with
suspect $r$-band photometry.

Fig.~\ref{compchi} shows the comparison of $\chi^2$ values for several
pairs of input models. Information on the $\mathrm H\alpha/\mathrm H\beta$
emission-line ratios is also shown since extinction turns out to be a
crucial parameter in the goodness of the model fits.  The shaded area
corresponds the zone of poor fits.  In the top-left diagram, BC99 and
\SB\, models with the same of the remaining parameters are compared.
Both models provide comparable results for most galaxies.

The bottom-left plot compares instantaneous and continuous star-formation
\SB\,  models. It is quite clear that better fits are obtained for most of the
galaxies with short bursts. A large fraction of the continuous
star-formation models are rejected by the observations.  There are a
handful of galaxies with better constant star-formation, but in all
cases almost equally good fits are obtained for the burst models.

The top-right diagram shows that the quality of the fits for
Miller-Scalo and Salpeter IMFs is indistinguishable. The same is true
for the Scalo IMF (not shown). At this point we are not able to
establish which of the tested IMFs best reproduces the observed
properties of the UCM galaxies.  We will return to this issue later.

Finally, the two extinction recipes are compared in the lower-right
panel. The \calz\, recipe seems to yield better fits than the
\cf\, one for high extinction objects (group of filled stars 
on the right). On the other hand, for some other galaxies, specially
those with low values of the $\mathrm H\alpha/\mathrm H\beta$ ratio,
\cf\, works better. For some cases neither provides confident results.

Fig.~\ref{bestone} shows the distribution of the best model fits for
the UCM galaxies according to the model input parameters. The $\chi^2$
estimator for each galaxy and model has been assumed to be the median
of all the 1000 Montecarlo particles and it has been normalized with
the number of colours used in its calculation.  For each galaxy, we
select the model that best-fittings its observational data, i.e
showing the lowest value of the $\chi^2$ estimator.

A total of 87 objects are best modelled with the \SB\, models rather
than with the BC99 ones.  This corresponds to 53\% of the complete
sample.  On average, these galaxies present redder observed $B-r$
colours and higher $EW(\mathrm H\alpha)$ values than the objects best
modelled with BC99 models: $(B-r)_{SB99}=0.9\pm0.3$
vs. $(B-r)_{BC99}=0.7\pm0.3$ and $EW(H\alpha_{SB99})=60\pm60$\,\AA\,
vs. $EW(H\alpha_{BC99})=110\pm90$\,\AA.  Moreover, the average
metallicity estimated by \SB\, models is lower than what BC99
predict. We will discuss these points in \pii.

We have only used \SB\, models with a Salpeter IMF. If we only
consider the galaxies best fitted with that IMF, the percentage of
best-fittings achieved with this evolutionary code increases to 73\%.

Fig.~\ref{bestone} also shows that 82\% of the UCM sample is best
described by an instantaneous burst of star formation. The objects
favouring a constant SFR are characterized by lower extinctions and
higher equivalent widths ($\langle E(B-V)\rangle=0.6$\,mag and
$\langle EW(\mathrm H\alpha)\rangle=168$\,\AA) than those best modelled
with instantaneous bursts (0.8 mag and 64 \AA).

Among the galaxies best modelled with the BC99 models, two of the IMFs
considered seem to dominate over the other one: the most common in
this distribution are the Salpeter IMF (42\%) and Miller-Scalo's (42\%
of the total number of galaxies best fitted by BC99 models). If we
also take into account the galaxies modelled with \SB\, templates, for
73\% of the galaxies a Salpeter IMF yields the best-fittings. These
results are in agreement with several studies claiming that a Salpeter
slope best reproduces the distribution of stellar masses in massive
star formation scenarios \citep[with perhaps a flattening at low
masses; see, for example,][]{1998ApJ...493..180M,1999A&A...347..532S,
2000A&A...354..802S,2000A&A...362...53S}. However, it is important to
emphasise that we have obtained these figures by a simple comparison
of the values of the $\chi^2$ estimator.  A proper discussion on the
IMF in UCM galaxies must involve parameters such as the upper mass
limit or the fraction of ionizing photons escaping from the birth
cloud. This is far beyond the scope of the present paper.

Finally, the \cf\, extinction recipe best reproduces the observed
colours and gas emission for 55\% of the sample. We notice again that
high extinctions prevail on the objects best fitted with the \calz\,
law, with $\langle E(B-V)\rangle=0.9\pm0.5$ (cf. $\langle
E(B-V)\rangle=0.6\pm0.4$ for \cf).

\begin{figure*}
\center{\psfig{file=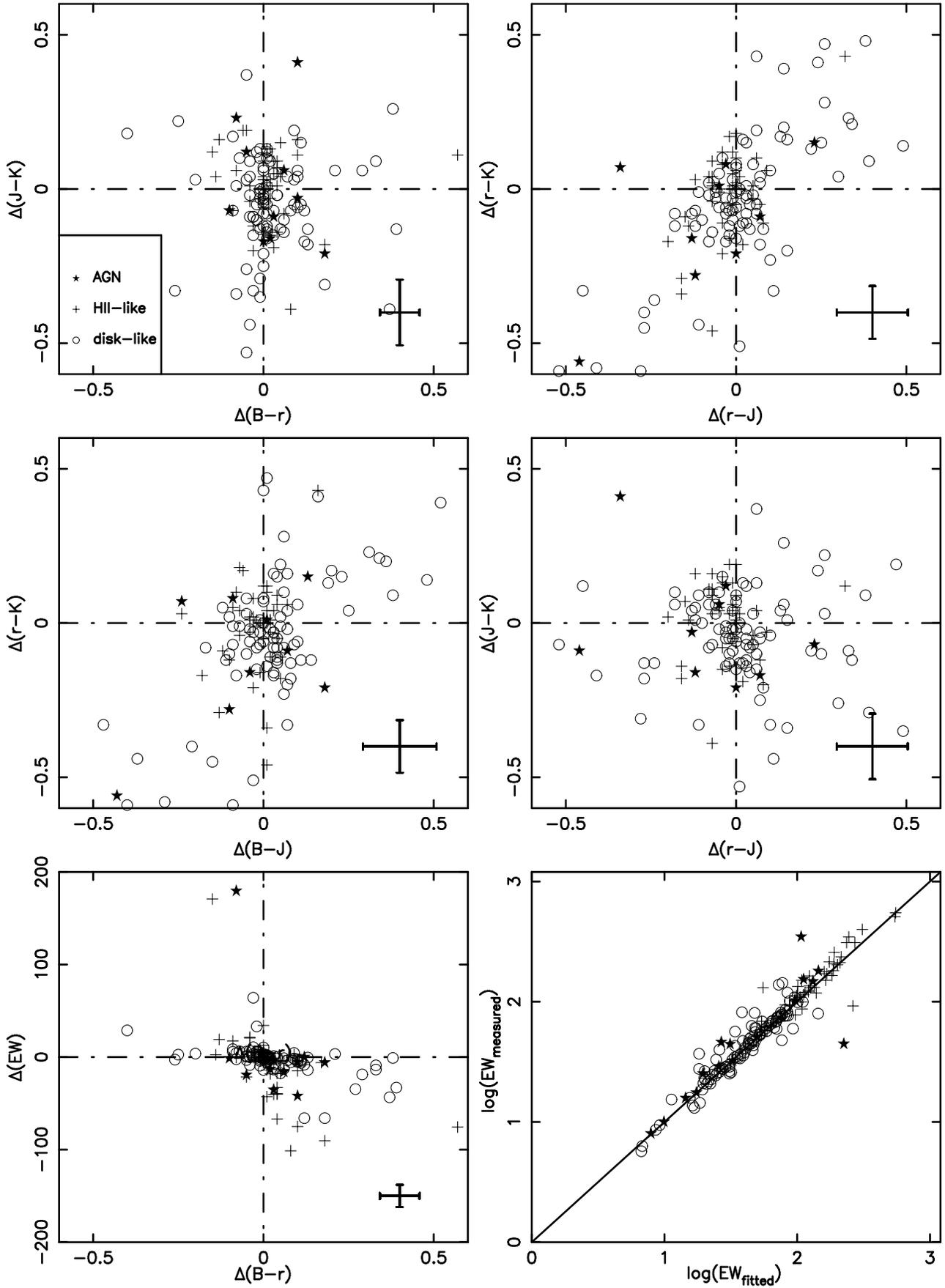,width=17cm}}
\caption{Differences between fitted and measured 
values for optical and nIR colours, and $EW(\mathrm H\alpha)$. Average
errors are shown in each panel. Different symbols stand for disk-{\it
like}, HII-{\it like} and AGN galaxies. The data refers to
instantaneous
\SB\, models with a Salpeter IMF and \cf\, extinction. }
\label{color1}
\end{figure*}

\begin{figure*}
\center{\psfig{file=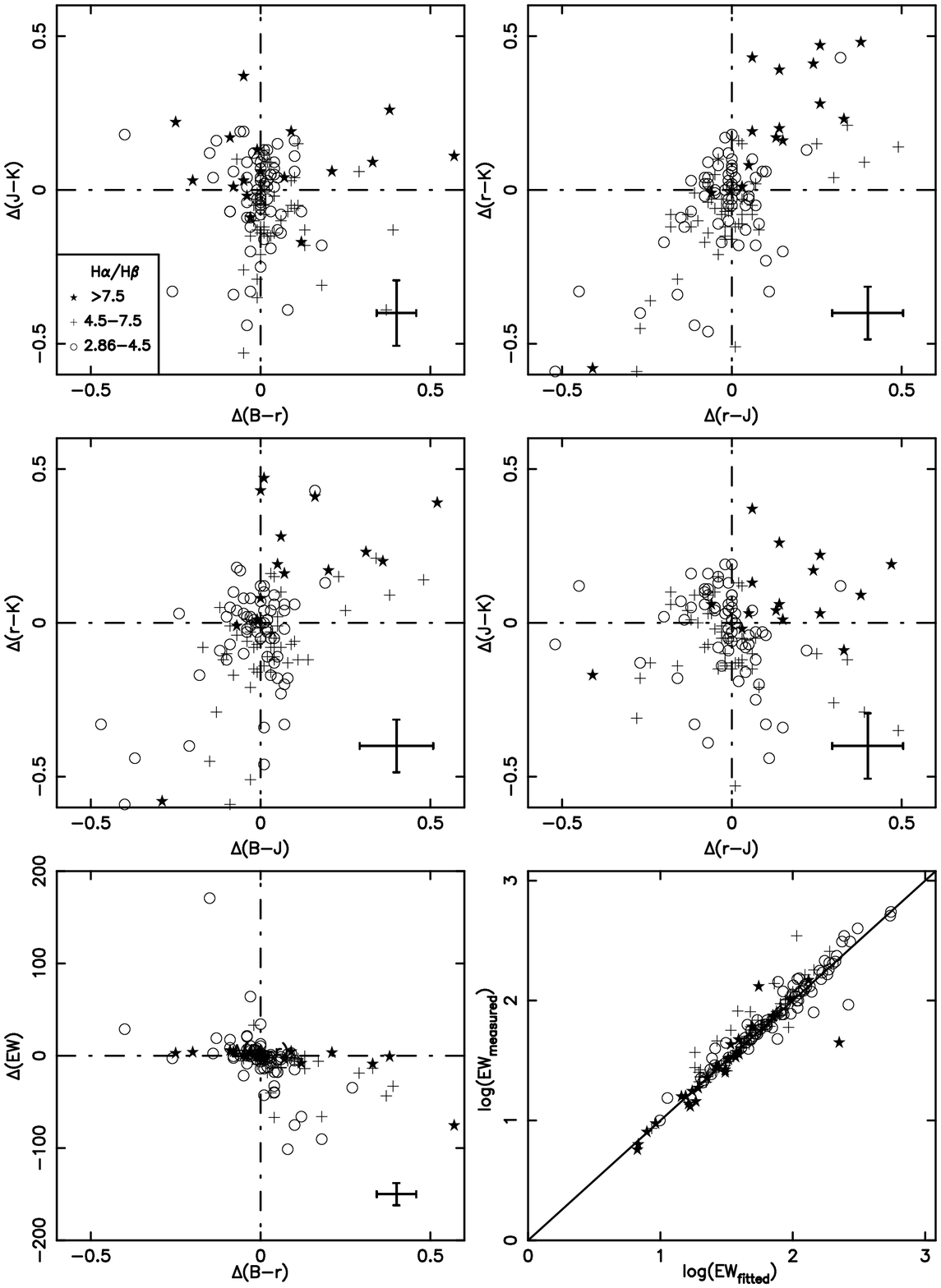,width=17cm}}
\caption{Same as in Fig.~\ref{color1} but the symbols represent different values
of the $\mathrm H\alpha/\mathrm H\beta$ ratio, an extinction indicator.}
\label{color2}
\end{figure*}

Figs.~\ref{color1} and~\ref{color2} present residual colour-colour
diagrams showing the differences between fitted and measured values
for several pairs of observables. Input parameters are \SB\, models,
instantaneous SFR, Salpeter IMF and \cf\, extinction
recipe. Information about spectroscopic type (Fig.~\ref{color1}) and
$\mathrm H\alpha/\mathrm H\beta$ ratio (Fig.~\ref{color2}) is also
shown in order to search for correlations between these quantities and
the goodness of the fit. The median error for each measured colour is
indicated by the error bars. In the case of $EW(\mathrm H\alpha)$ we
have plotted the lines of equality for fitted and measured values.

First, it is clear that the AGNs are not well-fitted (three other AGNs
are outside the boundaries of these plots, together with two of the
galaxies mentioned at the beginning of this section).  As expected,
the contribution of the active nucleus cannot be reproduced by the
stellar synthesis models. These AGN will be excluded from the rest of
the discussion.

A group of objects, mainly disk-{\it like} galaxies, exhibit a deficit
of observed $B$-band light: their $B-r$ and $B-J$ colours are redder
than the best-fit model predictions (e.g., objects with large
$\Delta(B-r)$ values in top-left panel). Most of these objects have
high $\mathrm H\alpha/\mathrm H\beta$ ratios. In some cases $\mathrm
H\beta$ was not detectable.  For the galaxies with undetected $\mathrm
H\beta$, an average $E(B-V)$ based on the spectroscopic type was used
initially, but this clearly underestimated the extinction and showed
fitted colours which were much bluer than the measured ones. For that
reason, we decided to use instead the average of the 25\% highest
$\mathrm H\alpha/\mathrm H\beta$ ratios for this spectral class. This
value was the one finally assumed and the one used to generate
Fig.s~\ref{color1} and~\ref{color2}. Although this yielded better
fits, it seems that we are still somewhat short of the real extinction
value for some objects.

At this point it is important to remind the reader that we are using
$EW(\mathrm H\alpha)$ and $\mathrm H\alpha/\mathrm H\beta$ values
measured in the long-slit spectra, and assume that they are
representative of the whole galaxy.

Another group of galaxies have optical-nIR colours which are not
well-fitted by the models, such as the object with positive
differences in the top-right panel. A visual inspection of these
objects reveals that a number of them are high-inclination galaxies
(ellipticity larger than 0.3), some with clear dust-lanes best
observed in the $B$ images. Examples include UCM0044$+$2246,
UCM2255$+$1654 and UCM2329$+$2427. The \cf\, extinction recipe fails
to model these highly-reddened galaxies (see Fig.~\ref{color2}), while
\calz\, provides better results. Among the 15 worst fitted objects of
this kind, 50\% have $EW(\mathrm H\alpha)$ lower than 30 \AA\, and
virtually all of the rest below 60
\AA. The observed $J-K$ colours for these galaxies are also redder than the
model predictions, indicating, perhaps, that the underlying old population is
more dominant in them. 

The problem with extinction gets obviously worse as we move to shorter
wavelengths. Some objects may be so extincted that we may be observing
just the `surface' of the galaxy disks in $B$ while we can see deeper
layers in the nIR \citep[see, for
example,][]{1996MNRAS.282.1005C}. Since we are observing fewer stars
in the blue bands, the measured colours would be redder than what the
models predict.  Moreover, significant uncertainties still remain in
the extinction recipes when trying to match observations spanning a
large wavelength range such as optical-nIR colours.

In the diagrams involving the $EW(\mathrm H\alpha)$ we see that the
models succeed reasonably well in fitting the observed data, although
there seems to be a relatively small tendency to underestimate the
observed values. Since the measured $\rm H\alpha$ $EW$s are based on
long-slit spectroscopy, and thus dominated by the central values, we
could be overestimating them if the star formation is significantly
more concentrated than the old stars.

\subsection{Solution degeneracy}

The technique that we have developed to derive the stellar properties
of the UCM galaxies is based on the use of the observational errors
and a Principal Component Analysis (PCA) study of the solutions. This
procedure allows us to obtain information about the degeneracy of the
results in the $\{t,b,Z\}$ parameter space.  In \gil\, we applied a
single linkage hierarchical clustering method
\citep{1987mda..book.....M} in order to study the clustering of
solutions achieved in the 1000 Montecarlo particles fitted for each
galaxy.  That paper pointed out that the clustering pattern is
dominated by the discretization in metallicity of the evolutionary
synthesis models. Thus, little can be learnt using this clustering
method before performing the PCA.  Instead, in the present work we
have applied the PCA to all the Montecarlo particles and obtained
average values and standard deviations for the entire set of solutions
of each galaxy.

\begin{figure}
\center{\psfig{file=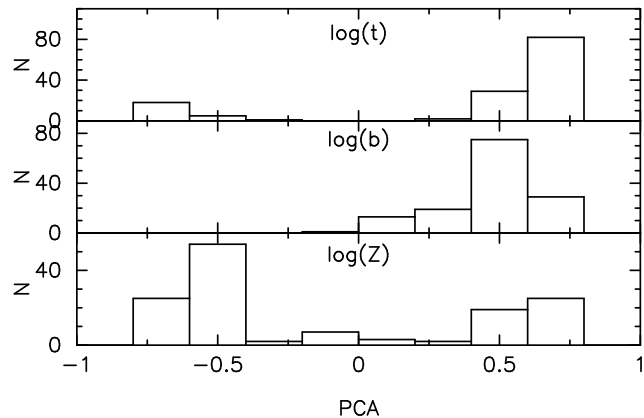}}
\caption{Histograms of the 3 components of the first vector of the PCA for
the UCM Survey galaxies. The plot refers to \SB\, models, Salpeter IMF,
instantaneous burst and \calz\, recipe.}
\label{pca}
\end{figure}

This method shows that, on average for the complete UCM sample,
$69\pm2$\% of the scatter of the Montecarlo particles is represented
by the first principal component in the PCA.  In less than 3\% of the
sample this fraction is less than half of the total scatter.  In
\gil\, the clustering characterization removed the scattering of the
solutions due to metallicity. The effect was that the component of the
PCA vector in the $Z$ direction was null in most cases.  Now the
distribution of this component for the whole sample is somewhat
flatter, with the strongest peak at $-0.5$ (see Fig.~\ref{pca}).  This
figure also shows that the age and burst strength components are
similar.  This means that both quantities are correlated: if we
increase the model age, we need to increase the burst strength in
order to keep the same $\mathrm H\alpha$ equivalent width.  Moreover,
since the strongest peak in the metallicity direction has opposite
sign to the other two, there is an age-metallicity degeneracy
(anti-correlation).

\section{Summary}
\label{summary} 

In this paper, the first of a series, we have described a method to
derive the properties of the star-formation and the stellar
populations in star-forming galaxies using broad-band photometry and
spectroscopy. We also present the available data for the UCM Survey
galaxies, covering the optical and nIR spectral ranges. The technique
is based on the assumption that our galaxies have a composite stellar
population. The evolved component resembles that of a typical
quiescent spiral/lenticular galaxy, whereas the young stellar
population component is generated with an evolutionary synthesis
model. This fact means that our modelling refers to the properties of
a recent star formation event which takes place {\it in excess of what
is typical in a normal spiral or lenticular galaxy}. The model
parameters considered are: (1) stellar evolutionary synthesis (the
Bruzual \& Charlot 1999 and \citealt{1999ApJS..123....3L} models); (2)
IMFs
\citep{1955ApJ...121..161S,1986FCPh...11....1S,1979ApJS...41..513M}; 
(3) star formation modes (instantaneous and constant); and (4)
extinction recipes (\citealt{2000ApJ...533..682C} and
\citealt{2000ApJ...539..718C}).

We have developed a statistical tool that takes into account the
observational uncertainties and a careful interpretation of the model
fits. The procedure is tested with the UCM sample data, and used to
study the dependence of the goodness of the model fits on several {\it
a priori\/} input parameters.  We find that our modelling is able to
reproduce the photometric and spectroscopic properties of almost all
the star-forming galaxies of the UCM Survey. Our test on the {\it a
priori} model parameter choices, based on our $\chi^2$ estimator,
reveals that:

\begin{itemize} 
\item[$\bullet$] both \SB\, and BC99 models provide reasonable and comparable 
fits.  The \SB\, models provide marginally better results, in
particular for redder galaxies with relatively higher $\mathrm H\alpha$
equivalent widths.
\item[$\bullet$] UCM galaxies clearly show a preference for 
instantaneous bursts of recent star formation rather than constant 
star-formation rates.
\item[$\bullet$] The models with a Salpeter initial mass function better 
reproduce the observations for nearly 75\% of the sample, although a
number of galaxies also present best results using the other IMFs and
this result must be regarded with caution.
\item[$\bullet$] The extinction description developed by \cf\,
yields satisfactory results for the majority of our sample galaxies
(with a variation in the extinction law), but it fails to reproduce
the properties of high extinction objects.
\end{itemize}

Among all the possible combinations of input parameters, an important
number of galaxies (one third) is best modelled with \SB\, code,
Salpeter IMF, instantaneous SFR and \cf\, extinction recipe.

In \pii, we will use the techniques developed here to study, in
detail, the properties of the UCM galaxies.

\section*{Acknowledgments}

This paper is partially based on data from CAHA, the German-Spanish
Astronomical Centre, Calar Alto, operated by the Max-Planck-Institute
for Astronomy, Heidelberg, jointly with the Spanish National
Commission for Astronomy. Also partially based on data obtained with
the 2.3m Bok Telescope of the University of Arizona on Kitt Peak
National Observatory, National Optical Astronomy Observatory, which is
operated by the Association of Universities for Research in Astronomy,
Inc. (AURA) under cooperative agreement with the National Science
Foundation. Also partially based on observations made with the Isaac
Newton and Jacobus Kapteyn Telescopes, operated on the island of La
Palma by the Isaac Newton Group in the Spanish Observatorio del Roque
de los Muchachos of the Instituto de Astrof\'{\i}sica de Canarias.

This research has made use of the NASA/IPAC Extragalactic Database
(NED) and the NASA/IPAC Infrared Science Archive which are operated by
the Jet Propulsion Laboratory, California Institute of Technology,
under contract with the National Aeronautics and Space
Administration. This publication makes use of data products from the
Two Micron All Sky Survey, which is a joint project of the University
of Massachusetts and the Infrared Processing and Analysis
Center/California Institute of Technology, funded by the National
Aeronautics and Space Administration and the National Science
Foundation.

PGPG wishes to acknowledge the Spanish Ministry of Education and
Culture for the reception of a {\it Formaci\'on de Profesorado
Universitario} fellowship. AGdP acknowledges financial support from
NASA through a Long Term Space Astrophysics grant to B.F.\
Madore. During the course of this work AAH has been supported by the
National Aeronautics and Space Administration grant NAG 5-3042 through
the University of Arizona and Contract 960785 through the Jet
Propulsion Laboratory. AAS acknowledges generous financial support
from the Royal Society. We also would like to thank George and Marcia
Rieke for kindly allowing us to use their near-infrared camera on the
University of Arizona 2.3m Bok Telescope. 

We are grateful to the anonymous referee for her/his helpful comments
and suggestions.

The present work was supported by the Spanish Programa Nacional de
Astronom\'{\i}a y Astrof\'{\i}sica under grant AYA2000-1790.

\label{lastpage}

\bibliographystyle{mn2e_pag}
\bibliography{referencias}

\end{document}